\documentclass[10pt,a4paper]{article}
\title{Kinematic quantities of finite elastic and plastic deformation}

\let\x\hskip        \let\y\vskip        \let\X\kern       
\let\xph\hphantom   \let\yph\vphantom   \let\ph\phantom
\let\bi\relax  \newdimen\bitlength  \bitlength=.06em      
\def\bit{   \x  \bitlength}  \def\biT{   \x -\bitlength}  
\def\bitt{  \x 2\bitlength}  \def\biTT{  \x-2\bitlength}  
\def\bittt{ \x 3\bitlength}  \def\biTTT{ \x-3\bitlength}
\def\bitttt{\x 4\bitlength}  \def\biTTTT{\x-4\bitlength}
\let\HB\hbox  \def\SB{\setbox1\HB}  \def\CB{\copy1}  
              \def\SC{\setbox2\HB}  \def\CC{\copy2}  
\def\RB#1{\raise#1\CB}  \def\XB{\wd1}  \def\YB{\ht1}  
\def\RC#1{\raise#1\CC}  \def\XC{\wd2}    
  
\let\sS\scriptstyle   \let\zS\scriptscriptstyle
\def\mathsizes#1#2#3{\mathchoice{#1}{#1}{#2}{#3}}
\def\textinmath#1{\mathsizes{\hbox{#1}}      
  {\hbox{\scriptsize #1}}{\hbox{\tiny #1}}}  
\def\txt#1#2#3{\x#1\bitlength\textinmath{#3}\x#2\bitlength}
\newcount\n  \newdimen\w

\usepackage{amsfonts}           

\def\mathbfit#1{{\textinmath{\textit{\textbf{#1}}}}} 
\def\mathbfitsf#1{{\textbfinmath{\textit{\textsf{#1}}}}} 
\def\textbfinmath#1{\mathsizes
  {\HB{\SB{#1}\def\P{\CB\X-\XB\RB{.03ex}}\P\X-\XB\X.07ex\P}}
  {\HB{\SB{\scriptsize #1}\def\P{\CB\X-\XB\RB{.03ex}}\P\X-\XB\X.07ex\P}} 
  {\HB{\SB{\tiny #1}\def\P{\CB\X-\XB\RB{.03ex}}\P\X-\XB\X.07ex\P}}}

\let\Cal\mathcal                       

\usepackage{amsmath}          

\def\told#1#2+{#1{#2}-}    
\def\az#1+{}

\def\Repeat#1#2{\n=#1\relax\loop\ifnum       
  \n>0\relax #2\advance\n by-1\repeat}

\long\def\OMIT#1{\relax }  

\def\typesize#1{\ifcase#1\tiny\or       
  \scriptsize\or\footnotesize\or\small\or\normalsize\or\large\or
  \Large\or\LARGE\or\huge\or\Huge\else\ifnum#1<0\tiny\else\Huge\fi\fi}
\newcount\currenttypesize     
\def\whichtypesize{\setbox0\hbox{m}\w=\ht0 \n=0 
   \loop\ifnum\n<10 \setbox0\hbox{\typesize     
   \n m}\ifdim\ht0=\w \currenttypesize=\n       
   \n=9 \fi \advance\n by1 \repeat}             
\newcount\documentptcase      
\def\whichdocumentptcase{\ifnum                
  \documentptcase=3 \setbox0                   
  \hbox{\small M}\w=.002\wd0                   
  \documentptcase=\w \divide\documentptcase    
  by 100 \advance\documentptcase by -11 \fi}   
\newcount           
  \currentmathsize  
\def\tenscptsize#1{\currentmathsize=\ifcase     
  #1 0\or0\or1\or1\or1\or2\or4\or5\or6\or8\fi}  
\def\elevenscptsize#1{\currentmathsize=\ifcase  
  #1 0\or0\or0\or0\or1\or1\or3\or5\or6\or8\fi}  
\def\twelvescptsize#1{\currentmathsize=\ifcase  
  #1 0\or0\or1\or1\or1\or2\or4\or5\or7\or7\fi}  
\def\tenscscsize#1{\currentmathsize=\ifcase     
  #1 0\or0\or0\or0\or0\or1\or1\or4\or5\or7\fi}  
\def\elevenscscsize#1{\currentmathsize\ifcase   
  #1 0\or0\or0\or0\or0\or0\or0\or3\or5\or7\fi}  
\def\twelvescscsize#1{\currentmathsize\ifcase   
  #1 0\or0\or0\or0\or0\or0\or2\or4\or6\or6\fi}  
\def\scptsize#1{\ifcase\documentptcase
  \tenscptsize{#1}\or\elevenscptsize{#1}\or\twelvescptsize{#1}\fi}
\def\scscsize#1{\ifcase\documentptcase
  \tenscscsize{#1}\or\elevenscscsize{#1}\or\twelvescscsize{#1}\fi}

\def\sect#1#2{{\section{#2}\label{#1}}}

\def\ssect#1#2{\subsection{#2}\label{#1}}

\long\def\quot#1{`#1'}  \long\def\quott#1{``#1''}

\long\def\emp#1{{\it #1}}  
\def\lat#1{{\it #1}}  

\def\eg{\lat{e.g.,\ }}   \def\ie{\lat{i.e.,\ }}

\def\re#1{(\ref{#1})}   
  
\def\eqn#1#2{ \begin{align} \label{#1}         #2 \end{align}}

\def\nl#1{          \\ \label{#1}        }  
\def\nnl#1{ \tag*{} \\ \label{#1}        }  
\def\nln#1{         \\ \label{#1} \tag*{}}  
\def\nnln#1{\tag*{} \\ \label{#1} \tag*{}}  
\def\m#1{$            #1         $}  
\def\mm#1{$   \,      #1  \,     $}  
\def\mmm#1{$  \,\,    #1  \,\,   $}  

\def\mat#1{ \begin{matrix}      #1 \end{matrix}}

\let\lt\left  \let\rt\right     
\def\p#1#2{\Delim#1{#2}\deliM#1}         
\def\pp#1#2{\lt\Delim#1{#2}\rt\deliM#1}  
\def\ppp#1#2#3{\delim                      
  #11#2{#3}\delim#12#2}                    
\def\delim#1#2#3{\csname\ifcase#1 relax\or   
   big\or Big\or bigg\or Bigg\fi\endcsname   
  {\ifcase#2\or\Delim#3\or\deliM#3\fi}}      
\def\Delim#1{\ifcase#1\relax\or(\or[\or   
  \{\or<\or\langle\or|\or\|\or---{ }\fi}
\def\deliM#1{\ifcase#1\relax\or)\or]\or   
  \}\or>\or\rangle\or|\or\|\or{ }---\fi}

\let\f\frac                     
\def\FF#1#2{(#1)/#2}        
\def\ff{\largerfrac{-1}}          
\def\fff{\largerfrac{-2}}        
\def\largerfrac#1#2#3{      
  \whichtypesize\n=\currenttypesize\advance\n by #1 \mathsizes
    {\SB{$   \vcenter{}$}\w=\YB\SB{\typesize\n$   \vcenter{}$}
      \advance\w by-\YB\raise\w\HB{\typesize\n$    \frac{#2}{#3} $}}
    {\SB{$\sS\vcenter{}$}\w=\YB\SB{\typesize\n$\sS\vcenter{}$}
      \advance\w by-\YB\raise\w\HB{\typesize\n$\sS{\frac{#2}{#3}}$}}
    {\SB{$\zS\vcenter{}$}\w=\YB\SB{\typesize\n$\zS\vcenter{}$}
      \advance\w by-\YB\raise\w\HB{\typesize\n$\zS{\frac{#2}{#3}}$}}}

\let\s\sqrt

\def\norm#1{\pp7{#1}}

\def\d{{\rm d}}        \let\pd\partial

\def\set#1#2{\xdef#1{#2}}                
\def\add#1#2{\n=#1\advance \n by #2\xdef#1{\the\n}}
\def\sub#1#2{\n=#1\advance \n by-#2\xdef#1{\the\n}}   
\def\mul#1#2{\n=#1\multiply\n by #2\xdef#1{\the\n}}
\def\div#1#2{\n=#1\divide  \n by #2\xdef#1{\the\n}}
\def\setadd#1#2#3{\set{#1}{#2}\add{#1}{#3}}
\def\setsub#1#2#3{\set{#1}{#2}\sub{#1}{#3}}
\def\setmul#1#2#3{\set{#1}{#2}\mul{#1}{#3}}
\def\setdiv#1#2#3{\set{#1}{#2}\div{#1}{#3}}
\def\addadd#1#2#3{\add{#1}{#2}\add{#1}{#3}}

\def\setaddadd#1#2#3#4{\setadd{#1}{#2}{#3}\add{#1}{#4}}

\def\setsubsub#1#2#3#4{\setsub{#1}{#2}{#3}\sub{#1}{#4}}

\def\setmuldiv#1#2#3#4{\setmul{#1}{#2}{#3}\div{#1}{#4}}
\def\setdivadd#1#2#3#4{\setdiv{#1}{#2}{#3}\add{#1}{#4}}

\let\bez\qbezier
\def\beztan#1#2#3#4#5#6#7#8{\setsub{\nP}{#5}{#1}\setsub{\nQ}{#6}{#2}%
\setmul{\nR}{\nP}{#8}\setmul{\nS}{\nQ}{#7}\setsub{\nP}{\nR}{\nS}\setmul
{\nR}{#3}{#8}\setmul{\nS}{#4}{#7}\setsub{\nQ}{\nR}{\nS}\setmul{\nR}{#3}
{\nP}\div{\nR}{\nQ}\setmul{\nS}{#4}{\nP}\div{\nS}{\nQ}\add{\nR}{#1}\add
{\nS}{#2}\bez(#1,#2)(\nR,\nS)(#5,#6)}

\def\nb{\makebox(0,0)}  

\def\insertright#1#2#3#4#5#6{\setbox0=\vtop{\null\noindent\resizebox
  {#2}{!}{\includegraphics{#6}}}\w=\columnwidth\advance\w by-#3\advance
  \w by-#2{\moveright \w\vbox to 0pt{\vskip#4\box0\vss}}\strut\advance
  \w by-#1\def\q{ }\n=1\loop\ifnum\n>#5\else\edef\q{ \q 0em \w }%
  \advance\n by1\repeat\parshape \n \q 0em \columnwidth}
\def\insertrightt#1#2#3#4#5#6#7{\setbox0=\vtop{\null\noindent\resizebox
  {#2}{!}{\includegraphics{#6}}\break #7}\w=\columnwidth\advance\w by
  -#3\advance\w by-#2{\moveright \w\vbox to 0pt{\vskip#4\box0\vss}}%
  \strut\advance\w by-#1\def\q{ }\n=1\loop\ifnum\n>#5\else\edef\q{ \q
  0em \w }\advance\n by1\repeat\parshape \n \q 0em \columnwidth}

\let\vect\mathbf      \let\tens\vect      
\let\Vect\Mathbf      \let\Tens\Vect      
  
\def\diad{\bitt\raisebox{.3ex}{\boldmath${\zS\otimes}$}\bitt}
\def\Per{/\X-.9ex/}  
\def\inv{^{-1}}
\def\tensUpRt#1{^{\rm\biT #1}}  
\def\tensUpRtIn#1#2{\tensUpRt{{#1}_{#2}}}  
\def\symm{\tensUpRt{S}}   
  \let\transp\trans
   \def\asymmm#1{\tensUpRtIn{A}{#1}}
\def\transs#1{\tensUpRtIn{T}{#1}}  
\def\tr{\txt00{tr}}

\def\dddot#1{\mathsizes
  {\SB{$#1$}\SC{$M$}#1\X-.5\XB\dot{\yph{#1}}\X-.2\XC
    \dot{\yph{#1}}\X.4\XC\dot{\yph{#1}}\X-.2\XC\X.5\XB}
  {\SB{$\sS #1$}\SC{$\sS M$}\dot{#1}\X-.5\XB\X-.2\XC\dot{\yph{#1}}\X
    .4\XC\dot{\yph{#1}}\X-.2\XC\X.5\XB}
  {\SB{$\zS #1$}\SC{$\zS M$}\dot{#1}\X-.5\XB\X-.2\XC\dot{\yph{#1}}\X
    .4\XC\dot{\yph{#1}}\X-.2\XC\X.5\XB}  }
\def\ddddot#1{\mathsizes
  {\SB{$#1$}\SC{$M$}\CB\X-.5\XB\X-.184\XC\ddot{\yph{#1}}\X
    .368\XC\ddot{\yph{#1}}\X-.184\XC\X.5\XB}
  {\SB{$\sS #1$}\SC{$\sS M$}\CB\X-.5\XB\X-.184\XC\ddot{\yph{#1}}\X
    .368\XC\ddot{\yph{#1}}\X-.184\XC\X.5\XB}
  {\SB{$\zS #1$}\SC{$\zS M$}\CB\X-.5\XB\X-.184\XC\ddot{\yph{#1}}\X
    .368\XC\ddot{\yph{#1}}\X-.184\XC\X.5\XB}  }

  \def\vess#1{{#1\kern.09em}'}

\def\Mater#1{\SB{$#1$}\CB\x-.91\XB      
  \widetilde{\hbox to .82\XB{\yph{$#1$}}}\x.09\XB}

\def\nablar{\nabla_{\x-.3ex{\zS\DDr}}}  \def\nablaR{\nabla^{}_{\biTT\DDR}}

\def\Dot#1{\overset{\mbox{\footnotesize \bf .}}{#1}}

\def\Dr{x}        \def\DDr{\vect{\Dr}}

\def\Dt{t}  \def\Dtnull{{\Dt_0}}    
            
\def\Du{u}         \def\DDu{\vect{\Du}}
\def\Dv{v}         \def\DDv{\vect{\Dv}}

\def\DC{C}        \def\DDC{\tens{\DC}}

\def\DE{E}        \def\DDE{\tens{\DE}}

\def\DH{F}      \def\DDH{\tens{\DH}}       
      
\def\DI{I}      \def\DDI{\tens{\DI}}

\def\DQ{O}          \def\DDQ{\tens{\DQ}}          

\def\DR{X}                 \def\DDR{\vect{\DR}}

\def\DU{U}  \def\DDU{\tens{\DU}}
\def\DDUL{\DDU_{\rm L}}      \def\DDUR{\DDU_{\rm R}}
\def\DDDUL{\DDU^{\yph{|}}_{\rm L}}  \def\DDDUR{\DDU^{\yph{|}}_{\rm R}}

\def\Dchi{\chi}  \def\DDchi{\Vect{\Dchi}}

\def\Aeps{\varepsilon}
  
\def\Aepsp#1{\ifcase#1\Aeps\para\or \dot{\Aeps}\para\or
  \ddot{\Aeps}\para\or \dddot{\Aeps}\para\or \ddddot{\Aeps}\para\fi}
\def\Aepso#1{\ifcase#1\Aeps\orth\or \dot{\Aeps}\orth\or
  \ddot{\Aeps}\orth\or \dddot{\Aeps}\orth\or \ddddot{\Aeps}\orth\fi}

\def\Asig{\sigma}  
  
\def\Asigp{\Asig\para} \def\Asigo{\Asig\orth}
\def\Asigp#1{\ifcase#1\Asig\para\or \dot{\Asig}\para\or
  \ddot{\Asig}\para\or \dddot{\Asig}\para\or \ddddot{\Asig}\para\fi}
\def\Asigo#1{\ifcase#1\Asig\orth\or \dot{\Asig}\orth\or
  \ddot{\Asig}\orth\or \dddot{\Asig}\orth\or \ddddot{\Asig}\orth\fi}

\def\LOmega{\Omega}  \def\LLOmega{\Tens{\LOmega}}

\def\paraline{\rule{.03em}{.9ex}}
\def\para{^{}_{\paraline\kern.17ex\paraline}}
\def\orth{^{}_{\biTT\perp}}

\def\four#1{\mathbfit{#1}}
\def\fourr#1{\mathbfitsf{#1}}  
\let\Vset\four
\def\Vpow#1{{\raisebox{.5ex}{$\sS #1$}}}
\def\VVVpow#1{\def\w{-.45ex}(\x\w(#1)\x\w)}
\def\VA{A}  \def\VVA{\four{\VA}}

\def\VC{C}        \def\VVVVC{\Cal{\VC}}
\def\VD{D}  \def\VVD{\four{\VD}}  \def\VVVD{\fourr{\VD}}
\def\VE{E}  \def\VVE{\four{\VE}}

\def\VI{I}  \def\VVI{\four{\VI}}
\def\VJ{J}  \def\VVJ{\four{\VJ}}
\def\VL{L}  \def\VVL{\four{\VL}}  \def\VVVL{\mathbb{\VL}}
\def\VM{M}  \def\VVM{\four{\VM}}  \def\VVVM{\mathbb{\VM}}
                                      \def\VVVVM{\Cal{\VM}}
  
\def\VR{R}  \def\VVR{\four{\VR}}  \def\VVVR{\mathbb{\VR}}
  
\def\VVRarb{\DDR^{\txt00{arb}}}
\def\VS{S}  \def\VVS{\Vset{\VS}}
\def\VU{U}  \def\VVU{\Vset{\VU}}      
\def\VV{V}  \def\VVV{\Vset{\VV}}      
\def\VW{W}  \def\VVW{\Vset{\VW}}  
\def\VT{T}    \def\VVVT{\mathbb{\VT}}
  
\def\Vd{d}  
  
\def\Vf{f}  \def\VVf{\four{\Vf}}
\def\Vg{g}  \def\VVg{\four{\Vg}}
\def\Vh{h}  \def\VVh{\four{\Vh}}
\def\Vk{k}  \def\VVk{\four{\Vk}}
\def\Vl{\ell}  \def\VVl{\hbox{\boldmath{$\Vl$}}}
\def\VVlen{\VVl}
\def\Vp{P}  
\def\Vq{q}  \def\VVq{\four{\Vq}}  \def\VVVq{Q}
\def\Vr{r}  \def\VVr{\four{\Vr}}
\def\Vu{u}  \def\VVu{\four{\Vu}}
\def\VVuC{\DDu^{\txt00{Cauchy}}}
\def\VVuarb{\DDu^{\txt00{arb}}}
\def\Vv{v}  \def\VVv{\four{\Vv}}
  
\def\Vt{t}  \def\VVt{\four{\Vt}}
\def\VGam{\Gamma}  \def\VVGam{\Tens{\VGam}}

\def\VVOmegaarb{\LLOmega^{\txt00{arb}}}
    
\def\Veta{\eta}  \def\VVeta{\Tens{\Veta}}
\def\Vtau{\tau}  \def\VVtau{\Tens{\Vtau}}

\def\tang#1#2{\four{T}^{}_{\!#1\!}\pp1{#2}}

\def\VVol{V}

\def\Fx{x}  \def\FFx{\four{x}}  \def\Fxa{\Fx_1}  \def\Fxb{\Fx_2}
\def\Fsig{\sigma}  \def\FFsig{\Tens{\Fsig}}

\def\Vo{o}

\begin{document}  

\author{{Tam\'as F\"ul\"op$^{1,2,3}$} and {P\'eter V\'an$^{1,2,3}$}\\
\\ \footnotesize  $^1$
Institute for Particle and Nuclear Physics, Wigner Research Centre for
Physics,
\\ \footnotesize  P.O.B. 49, H-1525 Budapest, Hungary 
\\ \footnotesize  $^2$ Dept.\ of Energy Engineering, Budapest Univ.\ of Technology and Economics,
\\ \footnotesize  Bertalan Lajos u.\ 4-6, H-1111 Budapest, Hungary
\\ \footnotesize  $^3$ Montavid Thermodynamic Research Group,
\\ \footnotesize  Igm\'andi u.\ 26.\ fsz.\ 4, H-1112 Budapest, Hungary}
\date{March 4, 2012}
\maketitle
 \begin{abstract}
Kinematic quantities for finite elastic and plastic deformations are
defined via an approach that does not rely on auxiliary elements like
reference frame and reference configuration, and that gives account of
the inertial-noninertial aspects explicitly. These features are achieved
by working on Galilean spacetime directly. The quantity expressing
elastic deformations is introduced according to its expected role: to
measure how different the current metric is from the relaxed/stressless
metric. Further, the plastic kinematic quantity is the change rate of
the stressless metric. The properties of both are analyzed, and their
relationship to frequently used elastic and plastic kinematic quantities
is discussed. One important result is that no objective elastic or
plastic quantities can be defined from deformation gradient.
 \end{abstract}

\sect{vab}{Introduction}

Processes of a continuum are determined by a set of equations, some
being balances \p1{of energy, momentum, etc.} and others being
constitutive relations among the various state-describing quantities of
the continuum. Among such quantities, some are related to the motion
of the medium. The usual way to describe the motion is to give it with
respect to a reference frame. This brings in an amount of subjectivity
and the question arises how it is possible to identify---or at least to
check---the frame independent physical content behind frame dependent
quantities.

A widely accepted textbook approach \p1{originally due to Noll
\cite{Nol67a,Nol58a}} prescribes a transformation rule for the
frame-related quantities under change of the reference frame to another
frame. Namely, the relationship between two frames is considered to
relate spatial and time coordinates \m{\bf x}, \m{t} recognized in one
frame to coordinates \m{{\bf x}'}, \m{t'} recognized by another frame
according to
 \begin{equation}
  t'=t,  \qquad\qquad
  {\bf x}'  ={\bf h}(t)+ {\bf Q}(t)({\bf x}-{\bf x}_0),
 \label{nollobj}\end{equation}
with a possibly time dependent rotation \( {\bf Q}(t)\) and
translation \m{{\bf h}(t)}, and an origin \m{{\bf x}_0}. Then,
if a physical quantity is transformed according to the Jacobian of this
transformation then it is considered objective \p1{carry a frame
independent physical meaning} in continuum physics.
However, this concept of objectivity is dubious because of two main
reasons:
 \begin{itemize}
\item It requires special, rigid reference frames.  
\item This transformation is actually four dimensional but continuum
mechanics utilizes only its three dimensional spatial part.
 \end{itemize}
The first problem has been pointed out many times and there are
suggestions to improve it \cite{Mar05a}. The second problem was
highlighted in \cite{MatVan06a,MatVan07a} and it was also shown how the
special definition \re{nollobj} hides the unavoidable nontrivial
intertwining of space and time, which is usually overlooked but actually
inherent also for Galilean spacetime \cite{Mat93b}.

Problematic aspects of objectivity and frame indifference are well-known
\cite{Fre09a,Mur03a,Mur05a,Mur83a,Spe87c,Rys85a,Liu03a,Liu05a,Liu09a,Cim11a,Mus98a,GurEta10b}.
Noll himself attempted to refine the solution several times. His last
works about the foundations of continuum mechanics aime at introducing a
frame-free formulation \cite{NolSeg10a}, starting from the concepts in
\cite{Nol67a}. However, spacetime remains implicit and time is separated
in his kinematics (see e.g in \cite{Nol06a}). In this respect, it is
similar to most of the above mentioned approaches, where the absolute
time creates the illusion of separability of spacetime into an absolute
space and an absolute time in Galilean physics.  

The usual indirect formulation of objectivity, with the help of
invariance properties under changes of reference frames, can be combined
with differential geometry for the spacelike part of the corresponding
physical quantities, too (see e.g.\ \cite{FabMar05a,Mar07a,YavEta06a}).
However, restricting ourselves to the transformation of reference frames
related to the spacelike part of physical quantities, thereby neglecting
their time dependence, results in an improper/insufficient formulation
of covariance, similarly to what was demonstrated for the case of the
rigid observers of Noll \cite{MatVan06a}. There are also several
approaches to the kinematics of elastic and plastic continua where a
Riemannian structure is emphasized for the purpose of elasticity, works
closely related to the questions raised here are
\cite{Hau02b,Ber05b,DiC05a,Eps99a,Leo99a,EpsElz07b}. Still, the
continuum is treated without its motion to be considered a process in
spacetime, a necessary step to obtain a frame independent theory.

Historically, the choice of a reference frame had usually been an
implicit step in theoretical descriptions. It was Einstein who made it
explicit, and emphasized the properties of observers, in his special
relativistic spacetime suggestion, which is in conform with Lorentz's
transformation rule between inertial observers, introduced in 1905. Some
years later came the next important achievement \cite{Wey18b,Wey22b},
\cite{Hav64a,Mat84b,Mat93b}, realizing that observers are not inevitable
to describe motion: any material point exists along a curve (world line)
in a four dimensional affine space (spacetime), and any observer is also
just a certain collection of material points, each with its world line
in spacetime (and organized together via a synchronization). As for the
spatial and temporal structure of spacetime, two variants are
well-known, the Galilean (also called nonrelativistic, Euclidean or
Newtonian) spacetime and the so-called special relativistic (also called
Minkowski) spacetime, the former being the frame free essence behind the
Galilean transformation rule of frames, and the latter being the version
behind the Lorentzian transformation rule. \p1{In addition, a non-affine
generalization of special relativistic spacetime is used for the general
relativistic theory of gravitation.}

Given the frame free physical notions and the corresponding mathematical
formalism, the program started to reformulate every frame dependent
quantity, equation and description. For example, it has been applied to
the mechanics of pointlike and of rigid bodies \cite{Mat86b}, to fluids
\cite{Mat86a} and to kinetic theory \cite{MatGru96a}.

In this paper, we continue this program for the question of how to
define elastic and plastic kinematic quantities in continuum physics
that are free from the dubious aspects of frame dependent descriptions.
Working on spacetime directly, without the need for observers, is an
automatically safe methodology to obtain a description that is frame
independent. Naturally, all results formulated without frames can
subsequently be translated (`projected') into frame dependent quantities
and expressions, for any frame chosen, which are the quantities directly
measured in laboratory and \textit{in situ} experiments.

First, the standard approach to kinematic quantities is reviewed for
further reference. This is followed by a survey of those questionable
aspects of these customary notions which motivate the subsequent
considerations. The fourth section describes the motion of a solid as a
process in spacetime. Here we introduce the two key notions, the current
and the relaxed metric, and then derive the kinematic quantities needed
for elasticity: elastic shape---the frame independent generalized
deformation---, and elastic deformedness---the frame independent
generalized strain. The resulting finite deformation compatibility
condition is calculated, too. The last section lays the foundation of
plastic kinematics. It defines the plastic kinematic quantity, which, in
any theory for plasticity, is to be determined via a constitutive law.
\p1{Such possible constitutive laws, derivable from various known
theories of plasticity, are not considered here.} The discussion closes
with a concise summary. The involved elements of the frame free
formalism of Galilean spacetime are collected in the Appendix.

Naturally, the most exciting aspect of the quantities introduced here is
to see how they work ``in action'', \textit{i.e.,} how one can build a
consistent continuum theory based on them. To establish such an example
can only be the matter of a separate full study. The authors mention
that construction of such a theory, combining elasticity, thermal
expansion, plasticity and rheology, is under good progress.

A word on terminology: in what follows, we use the term
\quot{kinematics} with the meaning \quot{branch of science that studies
how to describe the motion of bodies (objects)}. Whether or not forces
act on the bodies is irrelevant in this definition of kinematics, the
concern is what quantities are to be used to characterize the motion
itself.

\sect{vac}{A typical definition of kinematic quantities}

So as to set the context for expressing the motivations, we start
with summarizing a prototypical approach to elastic and plastic
quantities.

This starts with choosing a reference frame---this step being taken
only tacitly, usually. Next, the continuum is represented by a reference
configuration, i.e., by its location in the space of this frame at a
chosen reference instant \mm{ \Dtnull \bit.} More closely, each material
point is represented by its position \mm{\DDR} in this space at
\mm{\Dtnull \bit.} At time \mm{\Dt}, the position of a material point is
\mmm{ \DDr = \DDchi^{}_{\Dt}(\DDR) \bit,} and its velocity is \mm{ 
\DDv_{\Dt}(\DDR) = \dot{\DDchi}_{\Dt} (\DDR) \bit,} where overdot means
partial derivative with respect to time, \mm{\dot{\xph{.}} = \pd_\Dt
\bittt \big|_{\DDR}}. With the displacement since \mm{ \Dtnull \,,}
 \eqn{aab}{
  \DDu_{\Dt} (\DDR) := \DDchi_{\Dt} (\DDR) - \DDR \,,
 }
one also has
 \eqn{vfr}{
  \DDv = \dot{\DDu} \,.
 }

The so-called deformation gradient is introduced
as
 \eqn{vfl}{
  \DDH := \DDchi \diad {\nablaR} \,,
 }
with \m{\diad} denoting dyadic/tensorial product; in what follows, the
partial derivative operation \mm{\nablaR} will act to the left or to the
right depending on context, always to reflect the proper tensorial order
\p1{\quot{order of tensorial indices}}. The deformation gradient is
assumed to be invertible. This is ensured if one requires that \mm{ \det
\DDH } \p1{which is the Jacobian of the map \mm{ \DDchi }} is
nonzero, which physically means that the continuum can never be
singularly compressed.
 Following from its definition, \mm{ \DDH }
obeys the properties
 \eqn{vfm}{
  \dot{\DDH} = \DDv \diad \nablaR ,
 \qquad
  \dot{\DDH} \bit \DDH\inv = \DDv \diad \nablar ,
 \qquad
  \pp1{\DDH \diad \nablaR}\asymmm{2,3} = \tens0 \,;
 }
where in the middle formula velocity is considered in variables \mm{
\Dt, \DDr } \p2{the connection with the variables \mm{ \Dt, \DDR } being
established by \mm{\DDchi^{}_{\Dt}(\DDR) }}, and in the last formula
antisymmetrization is carried out in the second and third
\quot{indices}. Similarly, \mm{\symm} will stand for symmetric part,
\mm{\transp} for transpose and \mm{ \txt00{tr} } for trace, and, for
example, \mm{ \txt00{tr}_{1,3} } will denote contraction of the first
and third \quot{indices}.
 Note that, according to the chain rule of differentiation of composite
functions, a multiplication by \mm{ \DDH } from the right gives the
transition from the derivative \mm{ \nablar } of a quantity to the
derivative \mm{ \nablaR } \p1{and multiplication by \mm{ \DDH\inv } from
the right gives the opposite direction}.

Related to this, if one has a process from \mm{ \Dtnull } to \mm{ \Dt_2 }
then, for any \mm{ \Dt_1 } in between,
 \eqn{vfu}{
  \DDH_{\Dt_2}^{\p1{\Dt_0}} =
  \DDH_{\Dt_2}^{\p1{\Dt_1}} \bitt \DDH_{\Dt_1}^{\p1{\Dt_0}}
 }
where \mm{ \Dt_1 } is also used as another reference instant, and that's
why now the used reference instants are also displayed in superscript.
As a special case,
 \eqn{vfv}{
  \DDH_{\Dt_1}^{\p1{\Dt_0}} =
  \pp2{\DDH_{\Dt_0}^{\p1{\Dt_1}}}\inv \,.
 }

The deformation gradient \p8{as assumed to be invertible} admits a polar
decomposition:
 \eqn{vfn}{
 \DDH = \DDUL \bit \DDQ = \DDQ \bit \DDUR \,,
 }
with orthogonal \mm{\DDQ} and symmetric and positive definite \mm{\DDUL
= \s{\DDH \bit \DDH\transp},} \mm{\DDUR = \s{\DDH\transp \DDH} \,.}

The various deformation tensors (Cauchy-Green, Finger, \ldots) are
defined as various powers of \mm{ \DDUL } and \mm{ \DDUR }:
 \eqn{vfo}{
  \DDC_{\rm L}^{(n)}  &:=  \p1{\DDDUL}^\Vpow{n} \,,                   
 &                   
  \DDC_{\rm R}^{(n)}  &:=  \p1{\DDDUR}^\Vpow{n} .                   
 \qquad
  (n = \ldots, -2, -1, 0, 1, 2, \ldots).
 }
{}From them, the various strain tensors (Green-Lagrange, St.\ Venant,
Biot, Almansi, Hencky, \ldots) are derived, all expressing in some way
the deviation from the identity tensor \mm{ \DDI }:
 \eqn{aah}{                   
  \DDE_{\rm L}^{(n)}  &:=                   
 \f{1}{n} \ppp22{\DDC_{\rm L}^{(n)} - \DDI} =                   
 \f{1}{n} \ppp22{\p1{\DDDUL}^\Vpow{n} - \DDI} ,                   
 &                   
  \DDE_{\rm L}^{(0)}  &:=  \ln \DDDUL ,                   
 }                   
 \eqn{aai}{                   
  \DDE_{\rm R}^{(n)}  &:=                   
 \f{1}{n} \ppp22{\DDC_{\rm R}^{(n)} - \DDI} =                   
 \f{1}{n} \ppp22{\p1{\DDDUR}^\Vpow{n} - \DDI} ,                   
 &                   
  \DDE_{\rm R}^{(0)}  &:=  \ln \DDDUR                   
 }                   
\p2{the \mm{ n = 0 } cases \p1{Hencky strains} being the l'H\^ opital
limits of the \mm{ n \ne 0 } series}.
\p1{Naturally, not only \mm{f\p1{x} = \FF{x^n - 1}{n}} but more general
functional forms are also allowed.}
In addition, the Cauchy strain
\p8{with which each of the above strains coincides in the leading order
of \mm{ \DDH - \DDI }, i.e., for small strains} is defined as
 \eqn{vfq}{
  \DDE^{\rm Cauchy}_{} = \DDH\symm - \DDI =
  \pp1{ \DDchi \diad \nablaR }\symm - \DDI =
  \pp1{ \DDu \diad \nablaR }\symm\, .
 }

The important dynamical purpose with strain is to use it as the variable
on which elastic inner forces \p8{described by the elastic stress
tensor} as well as the corresponding elastic energy are assumed to
depend. When plastic changes also occur in a material, kinematics needs
to describe what a plastic change is geometrically, and how it differs
from elastic changes. One usual approach is to assume that strain
\p1{one of the definitions above, typically \mm{ \DDE^{\rm Cauchy}_{} }}
decomposes into a sum,
 \eqn{vfs}{
  \DDE^{\txt00{total}} = \DDE^{\txt00{elast}} + \DDE^{\txt00{plast}} \,,
 }
and another conception, considered more applicable for nonsmall
deformations, decomposes the deformation gradient, instead, and does it
multiplicatively:
 \eqn{vft}{
  \DDH^{\txt00{total}} = \DDH^{\txt00{elast}} \bitt \DDH^{\txt00{plast}} \,.
 }
The latter may be explained that, after a subsequent complete elastic
relaxation, which would bring in multiplication by \mm{
\p1{\DDH^{\txt00{elast}}}\inv } from the left \p2{in accord with
\re{vfu} and \re{vfv}}, one would obtain what the plastic deformation
is, and one would reach the relaxed configuration.

\sect{vad}{Remarks and observations}

The following remarks and observations will help in giving motivations
and hints for an improvement of kinematics.

\ssect{vbh}{So many strains}

Probably the most immediate observation is that there are many, actually
infinitely many definitions for deformation and for strain. In addition,
it is not hard to introduce infinitely many further \p8{and not
unreasonable} versions. Then, which one to use as the variable of an
elastic constitutive relation? In linear elasticity, in which strain the
elastic constitutive relation is expected to be the most linear? \p0{For
instance, Horgan and Murphy \cite{HorMur09a,BeaSta86a} finds that, for
large deformations of hard rubber, among the \mm{\DDE_{\rm R}^{(n)}}
strains, \mm{n = 0} provides the most precise linearity and in the
largest regime.} In parallel, in nonlinear elasticity, fitting
experimental data to a nonlinear elastic constitutive relation, e.g., to
the Murnaghan model, can provide unphysical values for the material
coefficients, and with large uncertainties, when, for example, the \mm{n
= 2} strain is used. When \mm{n = 0} is chosen, instead, then the
results are much more realistic and much more reliable \cite{PleKru06a}.

To summarize, there is no satisfactory amount of knowledge collected and
distributed about the relevance of the various strain measures.

\ssect{vbi}{Volumetric properties towards infinities and in between}

A less apparent, but not unimportant, aspect is that the strains
\re{aah}--\re{aai} with positive \mm{n} take finite value in the
infinitely compressed singular limit \p1{\mm{\det \DDH \to 0}}. On the
other side, the cases with negative \mm{ n } take finite value in the
infinitely expanded asymptotics \p1{\mm{\det \DDH \to \infty}}. A
geometrically really descriptive strain quantity would diverge in both
these singular limits. Also, it is such a quantity from which good
numerical stability could be expected. Namely, if, by numerical error,
the numerical representation of the system starts to deviate towards
large extension or compression, such a strain measures it sensitively,
and the proportional additional elastic forces will drive the situation
back towards the correct value.

In this respect, the Hencky strains do better than the others listed
above, as they diverge in both asymptotics. This, therefore, provides as
well a possible explanation of the findings of \cite{PleKru06a}
mentioned previously.

At this point, it is worth asking the following question, too. For small
deformations, the Cauchy strain is known to describe the purely
volumetric changes by its spherical part \p1{trace part}, and purely
torsional ones by its deviatoric part. Is there such a strain \p8{one
among those mentioned above, or some other one} that admits the same
property for finite deformations as well? Actually, one finds (see
\textit{e.g.,} \cite{Hol07b}) that the left and right Hencky strains
\p1{\mm{n = 0}} satisfy this criterion as well. Hence, this is another
aspect that distinguishes the Hencky strain.

\ssect{vbk}{Elastic kinematics is about a state, not a change}

Measuring strain with respect to a configuration at a \mm{\Dtnull} means
to measure a change, change that occurred during a time interval
\mm{\p2{\Dtnull, \Vt}}. However, this is not what we are physically
interested in. What we really have in mind is that we consider an
elastic solid body, which has a distinguished state, a natural state,
which the body takes when it's totally relaxed, undisturbed, with no
outer surface nor volume forces. Whenever this body is not in this state
then inner elastic forces rise that try to govern the body towards this
relaxed state. These forces somehow depend on how far the body is from
the relaxed state. In accord with that nonrelativistic
reversible/conservative interaction forces depend usually on the current
distance between the objects in interaction, we expect that the inner
elastic force \p1{elastic stress} depends on the current pairwise
distances between material points. More closely, the elastic stress
would depend on how the current distances differ from the relaxed
distances. We wish a geometric type state quantity
that compares, locally, the current pairwise distances with the relaxed
ones. To express its physical role, we can call it \emp{elastic
deformedness}, for example.

At an instant \mm{\Dtnull} chosen as an initial time, the elastic
deformedness of the body is, in general, nonzero. In many laboratory
testing situations, it can be considered zero---but not always: for
example, the Anelastic Strain Recovery method
\cite{MatTak93a,Mats08a,LinEta10p}
determines underground three dimensional \lat{in situ} stress by
measuring how rocks taken from the drill core relax from deformed to
undeformed state. In parallel, in a civil engineering underground
situation, initial deformedness of soil or rock is nonzero because of
self-weight and other effects.

The motion of the continuum known since an initial time determines only
how deformedness \emp{evolves} from the initial condition. For the
various different measures of deformedness, the rate equation expressing
this evolution is different.

It is instructive to derive what rate equation the deformations and
strains \re{vfo}--\re{vfq} obey. Actually, not all these rate equations
can be directly calculated because a symmetric tensor does not, in
general, commute with its time derivative. It is the left and the right
Cauchy-Green deformations and the Cauchy strain for which the
calculation can be done explicitly, finding
 \eqn{vfw}{
  \big({\DDUL^2}\Dot{\big)} & = \p1{ \DDv \diad \nablar } \DDUL^2 +
  \DDUL^2 \p1{ \DDv \diad \nablar }\transp
 \nl{vfx}
  \big({\DDUR^2}\Dot{\big)} &= 2 \bitt \DDH\transp
  \bi \p1{ \DDv \diad \nablar }\symm \bittt \DDH
 \nl{vfy}
  \dot{\DDE}^{\rm {Cauchy}}_{} &= \ppp11{\DDv \diad \nablaR}\symm \,.
}
New measures for deformedness can also be defined based on a rate
equation. For example, below we will make use of the \quot{inertial
version} of the Cauchy deformedness, characterized by the differential
equation
 \eqn{vfz}{
  \dot{\DDE}^{\rm {in. Cauchy}}_{} = \ppp11{\DDv \diad \nablar}\symm
}
\p2{with \mm{\nablar} rather than \mm{\nablaR} used in \re{vfy}}.

A consequence of the generally nonzero initial deformedness
at \mm{\Dtnull} is that
 \eqn{vga}{
  \DDE^{\rm Cauchy}_{} \not= \pp1{ \DDu \diad \nablaR }\symm
 }
in general. Only
 \eqn{vgb}{
  \DDE^{\rm Cauchy}_{\Dt} = \pp1{ \DDu_{\Dt} \diad \nablaR }\symm +
  \DDE^{\rm Cauchy}_{\Dtnull}
 }
holds.

Nevertheless, not only \mm{\dot{\DDE}^{\rm {Cauchy}}_{}} and
\p8{consequently} its integrals
 \eqn{vgd}{
  \DDE^{\rm Cauchy}_{\Dt} - \DDE^{\rm Cauchy}_{\Dtnull} =
  \int_{\Dtnull}^{\Dt} \dot{\DDE}^{\rm {Cauchy}}_{\Dt'} \d \Dt'
 }
satisfy the so-called Saint-Venant compatibility conditions \p2{which,
in Cartesian coordinates, is a consequence of Young's theorem of mixed
partial derivatives, applied on \re{vga}} but \mm{\DDE^{\rm
Cauchy}_{\Dt}} itself also:
 \eqn{vgc}{
  \nablaR \times \DDE^{\rm Cauchy}_{\Dt} \times \nablaR = \tens0 \,.
 }
Why \re{vgc} holds can be argued as follows. Zero deformedness does
naturally fulfil the compatibility condition. Furthermore, any nonzero
deformedness at a time \mm{\Dt} must be such that, by a subsequent
relaxation, the body could be brought into zero deformedness, and thus
being totally relaxed at a later \mm{\bar{\Dt}}. It is not required that
the actual process of the body is such that such a later zero
deformedness is reached indeed---but the possibility must hold for
that. Physically, we say here that a deformed elastic body must be able
to relax to undeformed state. Now, the integral
 \eqn{vhe}{
  \DDE^{\rm Cauchy}_{\bar{\Dt}} - \DDE^{\rm Cauchy}_{\Dt} =
  \int_{\Dt}^{\bar{\Dt}} \dot{\DDE}^{\rm {Cauchy}}_{\Dt'} \d \Dt'
 }
satisfies the compatibility condition and, for this hypothetical
\mm{\bar{\Dt}\,,} \mmm{\DDE^{\rm Cauchy}_{\bar{\Dt}} = \tens0} so
\mm{\DDE^{\rm Cauchy}_{\Dt}} also obeys the compatibility condition. An
analogous argument can be given for that \mm{{\DDE}^{\rm {in.
Cauchy}}_{\Dt}} fulfils the \mm{\nablar} version of the compatibility
condition. Note that later discussions will shed further light on this
\quot{ability to relax}.

At this point, it is important to recall the Ces\`aro--Volterra
formula,
 \eqn{lal}{
  \VVuC_{\Dt} (\DDR) & := \VVuarb_{\Dt}
  + \VVOmegaarb_{\Dt} \bi \ppp11{\DDR - \VVRarb}
 \nln{lam}
  &  \x2.7ex
  + \int_{\VVRarb}^{\DDR} \pp2{ \DDE^{\rm Cauchy}_{\Dt} (\DDR') +
  2 \p1{ \DDE^{\rm Cauchy}_{\Dt} \diad \nablaR }\asymmm{1,3} (\DDR')
  \bi \ppp11{\DDR - \DDR'} } \biTT \d \DDR' ,
 }
where, having the auxiliary position \m{\VVRarb} and the path of
integration fixed arbitrarily, \mm{\VVuC_{\Dt}} is uncertain up to the
arbitrary vector function \mm{\VVuarb_{\Dt}} and arbitrary
antisymmetric tensor function \mm{\VVOmegaarb_{\Dt}}---\ie
up to an arbitrary rigid body displacement-plus-rotation. The
Ces\`aro--Volterra formula produces all the possible vector functions
\mm{\VVuC} with which \mm{\DDE^{\rm Cauchy}_{}} satisfies
 \eqn{vge}{
  \DDE^{\rm Cauchy}_{\Dt} = \pp1{ \VVuC_{\Dt} \diad \nablaR }\symm \,.
 }
Such a \mm{\VVuC_{\Dt}} is, however, not the displacement
\mm{\DDu_{\Dt}} that is measured since a \mm{\Dtnull}, in general.
\mm{\DDu_{}} can be one of the possible \mm{\VVuC}'s only in the special
cases when initial deformedness is zero \p2{cf. \re{vgb}}, and even then
it is only one of the possible \mm{\VVuC}'s.

A \mm{\VVuC} is to be considered as a vectorial type potential for a
symmetric tensor that satisfies Saint-Venant's compatibility condition,
the condition to admit zero left-plus-right curl. Indeed, this situation
is an analogue of that a curl free vector field admits a scalar
potential, which is non-unique and can be obtained from the vector field
via its integral along an arbitrary curve. By this analogy, \mm{\VVuC}
can be named a \emp{Cauchy potential} of a Cauchy type tensor field
\p1{a symmetric tensor field with the Saint-Venant property}. This
naming explains the superscript \mm{{}^{\rm Cauchy}} in the notation
\mm{\VVuC\,.} A Cauchy potential is a mathematical auxiliary quantity,
which is in general unrelated to the physical displacement quantity: the
indeterminateness in \mm{\VVuC} allows displacement only as a special
case of \mm{\VVuC\,,} and even the possibility for this special choice
is ruined by nonzero initial deformedness. Cauchy potential and
displacement are only occasionally and weakly related quantities. These
two notions should never be confused.

For later use, let us give here a rewritten form of \re{vge},
 \eqn{vhv}{
  \DDE^{\rm Cauchy}_{\Dt} = \pp2{ \pp1{ \DDR + \VVuC_{\Dt} }
  \diad \nablaR - \DDI }\symm \,,
 }
too, despite that, at the moment, not any usefulness of this version is
apparent.

\ssect{vbl}{The aspect of the relaxed structure}

As has already been expounded, an elastic solid body possesses a relaxed
structure which tells all pairwise distances of material points in the
relaxed state, and elastic kinematics should express the deviation of
current distances from the relaxed ones. It turns out that this can be
done via a tensorial quantity \p1{this will be detailed in
Sect.~\ref{var} but is also plausible from the traditional approaches to
deformedness via strain}. For a liquid, we intend to do something else.
Namely, a liquid does not have distinguished distances and has
elasticity only volumetrically. There is a distinguished volume \p8{per
moles/particle-number/mass} which is realized for relaxed liquids
\p1{for amounts large enough so that surface tension can be neglected,
this volume is decided by volumetric elasticity only}. Elastic
force/stress in a nonrelaxed liquid depends only on how the current
volume differs from the relaxed one. Therefore, for liquids a scalar
quantity \p8{which could be called \emp{expandedness}, and could be
defined as
 \mmm{\p1{\VVol_{\Dt} - \VVol_{\txt00{relaxed}}}/{\VVol_{\txt00{relaxed}}}}
or its logarithm} is needed as the kinematic state variable for a
elastic constitutive relation.

In the case of a gas even less holds physically. Gases have infinite
relaxed volume so there only the current volume \p8{per
moles/particle-number/mass} remains as physically relevant kinematic
quantity.

Granular media, which might \p8{under certain circumstances} also be
described as a continuum, pose a further, nontrivial, question towards
kinematics.

To summarize, we can see that different types of continua require
different kinematic description, corresponding to the different physical
properties. This should be made explicit in any approach to continuum
kinematics, and the various types of media should be treated
differently.

One consequence of this observation is that {\it the deformation
gradient cannot be used for the definition of elastic deformedness}. The
deformation gradient quantity exists for hydrodynamic flows, too---it
could well be named \quott{flow gradient} as well---,
while elastic deformedness is physically meaningful only for solids.
Elastic deformedness is an additional state variable, existing in the
case of solids. In the \quott{differential equation plus initial
condition} picture shown above this is manifested in that the initial
condition for deformedness cannot be calculated, cannot be derived from
the deformation gradient at that initial time. The initial condition is
some independent quantity, and it is just the value of the elastic
deformedness state variable at that initial instant.

Hereafter, let us concentrate on solids only. There, having seen that
elastic kinematics is to compare the current distances to the relaxed
ones, it is plausible how plasticity is connected to this situation.
Indeed, plastic change means change of the relaxed structure, the change
of relaxed pairwise distances. Contrary to elastic kinematics,
plasticity is not about a \emp{state} but about a \emp{change}. None of
the relaxed structures is physically distinguished with respect to
another. For plastic kinematics, such a quantity is needed that tells
the rate of how the relaxed distances change.

This also implies that \re{vfs}, \ie \mmm{ \DDE^{\txt00{total}} =
\DDE^{\txt00{elast}} + \DDE^{\txt00{plast}} \,,} is problematic:
\mm{\DDE^{\txt00{elast}}} must mean elastic deformedness \emp{at a given
instant}, while \mm{\DDE^{\txt00{plast}}} can only be something like the
time integral of some rate of change of relaxed distances, between
\emp{two given instants}. The first term is a state and the second a
change, hence, their sum can be neither a state nor a change. These two
terms are conceptually different and physically incompatible. \p2{The
alternate formula of plastic kinematics, \re{vft}, will be commented in
the subsequent subsection.}

\ssect{vbm}{Arbitrary auxiliary elements in the description}

The above-described approach to continuum kinematics makes use of a
number of auxiliary elements, most of them being introduced for
technical convenience and formal simplification. These cause that the
description mixes the observed \p8{the phenomenon we wish to speak
about} with the observer---{us, who speak about the phenomenon}. Then we
need rules that tell what changes how when we modify the auxiliary
components so that the content on the phenomenon itself remains
unmodified. Experience shows that this is not so simple as we hope and
generates, in fact, controversies, like those seminal two which are
usually labelled as \quot{material frame indifference} and
\quot{material objectivity}.

Thus it is our practical interest to reveal all the auxiliary elements,
and then to devise such a description that is free from these
ingredients.
Let us therefore see now what these troublesome elements are in turn.

Probably the most eye-catching one is the choice of a reference instant
\mm{\Dtnull\,.} Time is homogeneous, no instant is more distinguished
than another so choosing one of them is an artificial step.

Most improved variants of continuum kinematics avoid this step and do
not assume elastic undeformedness at any time either, but use only a
reference configuration to describe undeformedness---{which
configuration may never be taken at any time during the lifetime of the
medium}. A reference configuration means to assign a location to each
material point. Unfortunately, the reference configuration still
contains arbitrary aspects. Both for elasticity and plasticity, the
spatial location and orientation of the body as a whole is irrelevant.
Space points are equivalent, and space directions are equivalent
\p1{space is homogeneous and isotropic}. Specifying them in some way
brings in arbitrary auxiliary elements in our description. Note that,
for deformedness, only pairwise material distances count \p1{the current
ones and the relaxed ones}.

For plastic kinematics, the not yet analyzed \re{vft}, \ie \mmm{
\DDH^{\txt00{total}} = \DDH^{\txt00{elast}} \bitt \DDH^{\txt00{plast}}
\,,} has been based on the following interpretation: after a complete
elastic relaxation \p2{multiplication by \mm{
\p1{\DDH^{\txt00{elast}}}\inv }} the plastic deformation \p8{embodying
the current relaxed structure} is reached. However, unfortunately, there
is no distinguished relaxed configuration as there is no distinguished
spatial location. Relaxation can occur in many different ways in space,
depending on that which material points of the body are kept fixed
during relaxation and which move, in what directions. As an example,
when we hold a sponge in our right hand and, with our left hand first
squeeze it and then release, then the end configuration depends on which
part of the sponge is kept fixed \p1{or moved} by our hand and where,
and in which direction we hold the sponge during its relaxing. We are
also allowed to move our right hand continuously and the sponge will
still relax---and move afterwards as a whole body.

A current relaxed structure includes the pairwise distances but not the
individual locations of material points. Consequently, the change of
relaxed structure \p8{plasticity} does not include locations or
velocities either. Both \mm{\DDH^{\txt00{elast}}} and
\mm{\DDH^{\txt00{plast}}} contain arbitrary aspects, they specify more
than is needed physically.

\ssect{vgf}{The aspect of reference frame}

At last, we need to examine what \quot{space} is. In other words, we
have to revisit the initial and usually implicit step of choosing a
reference frame, mentioned at the beginning of Section~\ref{vac}.

A reference frame is a collection of material points with respect to
which any pointlike object \p1{phenomenon} is observed by measuring its
distances from the reference material points at any time. A reference
frame is inertial if it sees free \p8{unaffected, uninfluenced} objects
to move in straight line with uniform velocity \p1{as a special case: to
be at rest with respect to the frame}. Inertial reference frames are a
special subclass of rigid frames, \ie of those frames where the pairwise
distances of reference material points are time independent. Nonrigid
frames also find applications.

In the approach to continuum kinematics considered in Section~\ref{vac},
it would be important to specify what reference frame is used. {}From a
noninertial frame, freely moving bodies are seen to have some
acceleration or rotation, while to be rest with respect to such a frame
requires some appropriate forces. Therefore, using a noninertial frame
brings in artifacts, the auxiliary element of using a reference frame
allows the presence of disturbing features which belong to the observer
and not to the phenomenon we wish to observe.
 Least \quot{harm} is done by using an inertial reference frame.

Galileo is known to point out first that inertial reference frames are
physically equivalent, neither of them is distinguished from the others.
His example to explain this was a ship on still water \cite{Gal632b}: if
we are in a windowless cabin within the ship, we cannot determine
whether the ship is at rest with respect to the shore or moves with some
uniform velocity to some direction, no matter what type of physical
experiment we perform in the cabin.

Inertial reference frames move uniformly with respect to each other.
When changing from one to another that moves with relative velocity
\mm{\mathbf{V}} with respect to the former, then time and position are
to be transformed as
 \eqn{aaq}{
 &&
  \Dt'  &=  \Dt ,
 &
  \DDr'  &=  \DDr - \mathbf{V} \Dt
 &&
 }
according to the Galilean transformation rule, which is applicable in
the nonrelativistic regime of relative velocities much smaller than the
speed of light.
 {}From this, we can read off that, Galilean relativistically, time is
absolute as the transformation rule for time is of the form
\mmm{\Dt' = f(\Dt)}. Similarly do we find that space is not absolute
since
\mmm{\DDr' \ne  \mathbf{g} \p1{\DDr}}.

{}From \mmm{\DDr' = \mathbf{g} \p1{\Dt, \DDr} = \DDr - \mathbf{V} \Dt} 
it is also clear that, to make space absolute, it must be combined with
time. Forming the four dimensional combination
\mm{ \zS  \pp1{ \begin{matrix} \Dt \\ \DDr \end{matrix} } ,
 }
the transformation rule reads
 \eqn{nah}{
  \pp1{ \begin{matrix} \Dt' \\ \DDr' \end{matrix} } =
  \pp1{ \begin{matrix} \Dt \\ \DDr -\mathbf{V} \Dt \end{matrix} } =
  \pp1{ \begin{matrix} 1 & \vect0 \\ -\mathbf{V} & \DDI \end{matrix} }
  \pp1{ \begin{matrix} \Dt \\ \DDr \end{matrix} }
 }
so the transformation does not lead out of the set of \mm{\zS \pp1{\mat{
\Dt \\ \DDr } }} values, and there is an absolute, physical, four
dimensional quantity \p1{four-vector} behind, whose representation is
\mm{\zS \pp1{\mat{ \Dt \\ \DDr } }} in one inertial frame and \mm{\zS
\pp1{\mat{ \Dt' \\ \DDr' } }} in another.

That space is not absolute is also clear from the fact that, for any
reference frame, the reference material points are the space points of
that frame, and they indicate the \quot{resting locations} as seen from
that frame. \p1{A side remark: in practice, for a frame we do not fill
the whole world with material points but take only a restricted amount
of them, and use interpolation-extrapolation rules for the other, empty,
locations.} Now, if a frame moves with respect to another then the
reference material points \p8{the resting locations} do not rest with
respect to the other frame. What is a space point for one frame is a
process for the other one.

This actually coincides with our everyday experience as well. When
travelling on a train, our reference frame is fixed to the train: the
seats, the window, etc. Even when our drink spills out of our cup
because of a fast breaking of the train, we view this from a viewpoint
fixed to the train---although the drink wished to continue its
inertial motion and it was the train that did something noninertial. In
the meantime, someone seeing this little accident from the outside,
standing on the earth nearby the rail, sees us, the drink and the whole
train as moving.

There is not one space as \quot{the universal background} of motions.
There is one spacetime which is the universal background of motions,
while spaces are relative, they are relative to frames/observers.

Certainly, to take the step of combining space with time has not been
easy historically, both because of physical and mathematical aspects.
Physically, time is experienced to be different from space in some
senses. In parallel, mathematically, geometry originally meant a
Euclidean space, and it took much time to gradually realize that
geometries other than Euclidean are also conceptually acceptable---and
even relevant. 

The four dimensional combinations show not a four dimensional Euclidean
structure but the combination of two other structures. One of them is
the linear map that takes the time component \mm{\Dt} out of \mm{\zS
\pp1{\mat{\Dt \\ \DDr}}}, and the other is a three dimensional Euclidean
scalar product
 \eqn{vgk}{
  \pp1{ \begin{matrix} 0 \\ \DDr_1 \end{matrix} } \cdot
  \pp1{ \begin{matrix} 0 \\ \DDr_2 \end{matrix} } =
  \DDr_1 \cdot \DDr_2
 }
for the spacelike combinations \mm{\zS \pp1{\mat{0 \\ \DDr}}} \p8{the
ones which have zero timelike component}, which combinations are
actually those invariant under Galilean transformation:
 \eqn{vgj}{
  \pp1{ \begin{matrix} 0 \\ \DDr \end{matrix} }' =
  \pp1{ \begin{matrix} 0 \\ \DDr - \mathbf{V} \cdot 0 \end{matrix} } =
  \pp1{ \begin{matrix} 0 \\ \DDr \end{matrix} } .
 }
The time component is absolute, and consequently simultaneity of
spacetime points also, and spacelike four-vectors with their scalar
product are also absolute.

Historically, only the advent of special relativity helped to rethink
the Galilean relativistic space and time ideas as well. That's why Newton
rejected Galileo's result on the equivalence of inertial reference
frames, and claimed the existence of an absolute space \p1{a
distinguished frame}: he could not imagine any other background for his
action-at-a-distance forces. He didn't know that absolute simultaneity
plus absolute spacetime vectors with Euclidean structure would suffice
for this purpose, and, at the same time, would allow to incorporate the
equivalence of inertial reference frames, too.

Some years after the birth of special relativity, another important
achievement took place: Weyl published such a formalism for spacetime
\p8{he established both the Galilean relativistic and the special
relativistic version} that does not rely on reference frames
\cite{Wey18b,Wey22b}. Spacetime points and their relationships, \ie the
structures of spacetime, are defined first, then pointlike objects are
modelled with world lines in spacetime, and reference frames are
introduced only later, as certain extended objects moving in spacetime.

This approach has gained attention only gradually, and was popularized
by Arnold, for example \cite{Arn78b}. Later, the quite succinct and basic
treatment by Weyl was expounded, enriched and worked out in detail for
the purposes of physics by Matolcsi \cite{Mat84b,Mat93b}.

The frame free spacetime formalism enables us to remove the reference
frame, the last arbitrary auxiliary element listed in this
Section, from continuum kinematics. Material objectivity can be viewed
directly, without the disturbing technical presence of frames. Also, the
physical intention behind the question of material frame indifference can be
clarified by rewording it as the question of noninertial motion
indifference \p1{by which we mean the question ``does some interaction
within the material couple to the acceleration or the angular
velocity of the continuum?''}.
 While the present work does not intend to contribute directly to the
topic of noninertial motion indifference, that question will also be
possible to investigate on the grounds layed down here in the subsequent
Sections: a frame free discussion of kinematics, with all the other
auxiliary ingredients also avoided.

Hereafter, let us follow the general methodology of mathematical models
for physical phenomena \cite{Mat86b}, with all notions introduced as
mathematical objects, defined fully mathematically, so as not to leave
any property obscured or tacitly assumed.

\sect{var}{The material manifold of solids}

Let the mathematical model for a continuum be a three dimensional simply
connected complete smooth manifold \mm{\VVVVC \,.} Its tangent space at
a material point, an element \mm{\Vp} of \mm{\VVVVC}, is
\mm{\tang{\Vp}{\VVVVC} \,.} The differential map of a smooth mapping
\mm{\phi} from \mm{\VVVVC} to an arbitrary smooth manifold \mm{\VVVVM}
at a point \mm{\Vp} is a linear map \mm{ \tang{\Vp}{\VVVVC} \to
\tang{\phi\p1{\Vp}}{\VVVVM} \,,} denoted by \mm{\p1{\phi \diad
\Mater\nabla} \p1{\Vp}\,.}
 Henceforth, overtilde will indicate material vectors, covectors, tensors
etc., \ie objects related to the tangent spaces of \mm{\VVVVC} \p1{\ie
referring to material directions} to make a distinction to vectors etc.\
related to the space \mm{\VVS} of spacelike spacetime vectors \p8{spatial
directions} \p1{as well as to quantities related to the spacetime vector
space \mm{\VVM}}. For notations, terms and other utilized elements of
the frame free formalism of Galilean spacetime, see the Appendix.

A motion or spacetime process of such a body is given by assigning a
world line function \mm{\Vr} to each material point \mm{\Vp} in a smooth
enough way. Thence, at a time \mm{\Vt\,,} point \mm{\Vp} is at spacetime
point \mmm{\Vr_\Vt\p1{\Vp}\,.} For a fixed \mm{\Vt\,,} \mmm{\Vp \mapsto
\Vr_{\Vt}\p1{\Vp}} is a mapping from \mm{\VVVVC} to \mm{\Vt} \p8{let's
recall that a \mm{\Vt} is a Euclidean affine space, a three dimensional
\quot{slice} within spacetime}, with its differential map or Jacobian
map at \mm{\Vp}
 \eqn{vgv}{
  \VVJ\biT_{\Vt}\p1{\Vp} = \p1{\Vr_\Vt \diad \Mater\nabla} \p1{\Vp}\,.
 }
This Jacobian is the frame free, reference time free and reference
configuration free generalization of the deformation gradient.
 It can be named \textit{world line gradient}.

The four-velocity of a material point \mm{\Vp} at a time \mm{\Vt} is
 \eqn{vgw}{
  \VVv_\Vt \p1{\Vp} = \dot{\Vr}_\Vt \p1{\Vp} \,.
 }
For convenience, hereafter let us not distinguish in notation functions
defined on \mm{\VVVVC} \p1{Lagrangian description; \eg \mm{\VVv_\Vt}}
from their Eulerian-like spacetime counterpart \p1{their composition
with \mm{\Vr_\Vt^{-1}} for each \mm{\Vt\,;} \eg \mm{\VVv_\Vt^{} \circ
\Vr_\Vt^{-1}}}. It is easy to check that, for a function \mm{f} defined
on spacetime, the comoving derivative \mm{\dot f} is to be introduced as
\mmm{\p1{f \diad \VVVD} \VVv} so that it be in accord with the overdot of
Lagrangian variabled functions.

Using the chain rule of differentiation of composite functions,
 \eqn{vgx}{
  \VVL_\Vt := \p0{ \p0{ \VVv_\Vt^{} } \diad \nabla }
 }
can be expressed with \mm{\VVJ_\Vt} as
 \eqn{vgy}{
  \VVL_\Vt^{} =
  \dot{\VVJ\bitt}\biTTTT_\Vt^{} \bitt \VVJ\biTT_\Vt\inv \,.
 }

At any instant \mm{\Vt\,,} the current distance of two material points
\mm{\Vp, \VVVq} is the distance of the two spacetime points
\mm{\Vr_\Vt\p1{\Vp}, \Vr_\Vt\p1{\VVVq}} where the two material points
stay at \mm{\Vt \,,}
 \eqn{vgz}{
  \Vd_\Vt \p1{\Vp, \VVVq} = \norm{ \Vr_\Vt \p1{\VVVq} - \Vr_\Vt \p1{\Vp}
  }_\VVh .
 }
This induces a Riemann metric tensor on \mm{\VVVVC}, which proves to be
 \eqn{vha}{
  \Mater{\VVh}_\Vt^{} = \VVJ_{\biTT\Vt}^* \VVh \bitt \VVJ_{\biTT\Vt}^{}
  \,,
 \qquad
  \Mater{\VVh}_\Vt^{} \p1{\Vp} : \tang{\Vp}{\VVVVC} \times
  \tang{\Vp}{\VVVVC} \to \VVVL^{\VVVpow{2}} \,.
 }
Actually, the following three characterizations of distance
relationships are equivalent on a Riemann manifold: 1) lengths of
curves, 2) distances of pairs of points \p1{defined as the length of the
shortest, geodesic, curve between the two points}, 3) the metric tensor.
The third choice uses a local quantity while the other two are nonlocal
characterizations so it is favourable to work in terms of the metric
tensor.

So far we haven't restricted our discussion to elastic solids. To make
this step, let us realize the physical ideas put forward in
Sects.~\ref{vbk} and \ref{vgf} by adding to our model of the continuum a
mathematical object that expresses the relaxed/stressless distances of
pairs of material points. This means that we furnish our material
manifold with a relaxed Riemann metric tensor \mm{\Mater\VVg\,.} In
general, this metric is different from the currently realized metric
\mm{\Mater{\VVh}_\Vt}. In other words, in general the body is not
relaxed, it is not undeformed, not even within some local neighbourhood.
It might be considered as a \quot{blessed moment} when the two metrics
coincide.

Nevertheless, we require something from the relaxed metric: that, under
appropriate circumstances, the current metric \emp{could} evolve into
the relaxed metric: that it is possible to bring the body into relaxed
state, even within a finite time interval, by some finite later time
\mm{\bar\Vt}, via some motion \mm{\bar\Vr} that would connect smoothly
to the current motion \mm{\Vr_\Vt} of the material and would continue in
such a way that the distances realized at \mm{\bar\Vt} coincide with the
relaxed ones for the whole body. \p1{In short: The relaxed metric must
be realizable as a current metric at some time \mm{\bar\Vt}. A solid
must have the ability to relax.} In the language of metrics, this is
expressed as \mmm{\Mater{\VVh}_{\bar\Vt} = \Mater\VVg\,.}
 Writing this in terms of \mm{\bar\Vr_{\bar\Vt}} and its gradient
\mmm{\bar{\VVJ\bittt}_{\biTTTT\bar\Vt} = \bar\Vr_{\bar\Vt} \bitt \diad
\nabla} reads
 \eqn{vhb}{
  \bar{\VVJ\bittt}_{\biTTTT\bar\Vt}^* \VVh \bitt
  \bar{\VVJ\bittt}_{\biTTTT\bar\Vt}^{} = \Mater\VVg\,,
 }
which tells that \mm{\bar\Vr_{\bar\Vt}} establishes an isometry, a
metric preserving diffeomorphism, between \mm{\VVVVC} as a Riemann
manifold with metric \mm{\Mater\VVg} and \mm{\Vt} as a Riemann manifold
with metric \mm{\VVh\,.}

Now, we know that \mm{\Vt} is actually a Euclidean affine space and,
hence, a flat Riemann manifold \p1{\ie having zero Riemann curvature
tensor}. Due to a theorem (see eg. \cite{Die60b}), a Riemann
manifold is isometric to a flat one iff it is flat. Therefore, the
relaxed structure on \mm{\VVVVC} is required to be a flat Riemann metric.

This condition can be further reformulated in a technically more
convenient way. Namely, \mm{\VVVVC} being three dimensional, not only its
Riemann tensor determines its Ricci tensor but the Ricci tensor also
determines the Riemann tensor \p1{one proof of this can be based on p88
of \cite{ONe83b}}. Consequently, the requirement of flatness is
equivalent to the zeroness of the Ricci tensor. To summarize, the
physical expectations dictate the relaxed structure of a solid body to
be a Ricci-flat Riemann metric.

\sect{vhc}{Elastic kinematics}

\ssect{vhh}{Elastic shape}

In the light of Sect.~\ref{vbk}, elastic kinematics needs a quantity
that expresses how the current distances \p8{on \mm{\VVVVC}, the current
metric \mm{\Mater{\VVh}_\Vt\,,}} differ from the relaxed distances ---
{on \mm{\VVVVC}, from the relaxed metric \mm{\Mater\VVg}}. Or, in terms
of spatial spacetime tensors, the current distances \p8{on \mm{\Vt},
\mm{\VVh}} are to be compared to the relaxed distances --- on \mm{\Vt},
to
 \eqn{vhf}{
  \VVg_\Vt := \pp1{\VVJ_{\biTT\Vt}\inv}^* \Mater\VVg \bittt
  \VVJ_{\biTT\Vt}\inv \,.
 }
Both \mm{\Mater{\VVh}_\Vt} and \mm{\Mater\VVg} are
\mmm{ \tang{\Vp}{\VVVVC} \to \pp2{ \tang{\Vp}{\VVVVC} \Per
\VVVL^{\VVVpow{2}} }^*}
type linear maps, and both \mm{\VVh} and
\mm{\VVg_\Vt} are
\mmm{ \VVS \to \pp2{\VVS \Per \VVVL^{\VVVpow{2}}}^*}
type maps.

It is also instructive to remark that it would be advantageous if the
elastic kinematic quantity agreed with the Cauchy strain in the regime
of small deformedness. We can formulate its definition \re{vfz} in the
frame free approach via its evolution equation as
 \eqn{vhd}{
  \dot{\VVE}{}^{\bit\hbox{\scriptsize in.\,{Cauchy}}}_{} &= \ff{1}{2}
  \ppp11{ \VVL + \VVh\inv \VVL^* \VVh } = \ff{1}{2} \ppp11{ \VVL +
  \VVL^+ } = \VVL\symm  \,,
 }
where the appearance of \mm{\VVh} is unavoidable and the
\mm{\VVh}-adjoint
 \eqn{vhg}{
  \VVL^+ := \VVh\inv \VVL^* \VVh
 }
of \mm{\VVL = \VVv \diad \nabla} must be used because \mm{\VVL} is an
\mm{\VVS \to \VVS \Per \VVVT} tensor, and its dual transpose, a
\mm{\VVS^* \to \VVS^* \Per \VVVT} tensor cannot be added to it, except
when \mmm{ \VVh : \VVS \to \ppp11{\VVS \Per \VVVL^{\VVVpow{2}}}^* } is
also inserted as appropriate. The resulting
\mm{{\VVE}{}^{\bit\hbox{\scriptsize in.\,{Cauchy}}}} is \mm{\VVS
\to \VVS} valued, and is \mm{\VVh}-symmetric.

Incidentally, the reinterpretation of the traditional Cauchy strain
\mm{\DDE^{\rm Cauchy}} would require to involve the Jacobian \mm{\VVJ}
as well, since the material gradient of velocity maps from
\mm{\tang{\Vp}{\VVVVC}} to \mm{\VVS} so it can be combined with its dual
transpose only when an appropriate map between \mm{\tang{\Vp}{\VVVVC}}
and \mm{\VVS} is also made use of.
 Note that the Jacobian tensor \mm{\VVJ} \p1{the spacetime consistent
generalization of the deformation gradient} can never be \quot{close to
the identity tensor} since it connects \emp{different} vector spaces so
the traditional Cauchy strain would become a rather complicated object
when seen in the frame free, reference time free and reference
configuration free way. In what follows, let us use \p1{for comparison
purposes and for checking the small deformedness regime} only the
inertial version \mm{{\VVE}{}^{\bit\hbox{\scriptsize
in.\,{Cauchy}}}_{}}, which is a simple and clear object.

Now, having seen that from which vector space and to which vector space
the current metric tensor and the relaxed metric tensor map, mathematics
allows to compose these two linear maps only in the combination
 \eqn{vcd}{
  {\VVA}_{\bit\Vt}^{} := \VVg_{\bit\Vt}\inv \bit \VVh
 }
as the spacelike spacetime tensorial version, and as
 \eqn{ver}{
  \Mater{\VVA}_{\bit\Vt}^{} = \VVJ_{\biT\Vt}\inv {\VVA}_{\bit\Vt}^{}
  \bitt \VVJ_{\biT\Vt}^{} = \Mater{\VVg}\inv \bitt
  \Mater{\VVh}_{\bit\Vt}^{} \,,
 }
the material equivalent of \mm{\VVA_{\bit\Vt}}. It is easy to check that
\mm{\VVA_{\bit\Vt}} is an \mm{\VVh}-symmetric \mm{\VVS \to \VVS} tensor,
while \mm{\Mater{\VVA}_{\bit\Vt}} is a \mm{\Mater{\VVg}}-symmetric
\mm{\tang{\Vp}{\VVVVC} \to \tang{\Vp}{\VVVVC}} type tensor.

If the material is in its relaxed state, \mm{\Mater{\VVh}_{\bit\Vt} =
\Mater{\VVg} \,,} \mm{\VVg_{\bit\Vt} = \VVh\,,} then
\mm{\Mater{\VVA}_{\bit\Vt}\p1{\Vp} = \VVI_{\biT\tang{\Vp}{\VVVVC}}\,,}
\mm{{\VVA}_{\bit\Vt} = \VVI_{\biT\VVS} \,.} This seems to indicate that
we have introduced not a deformedness \p1{a reinterpreted strain}
but a reinterpreted deformation. We can call it the \emp{elastic shape
tensor}.

It is possible to derive the time evolution equation of the elastic
shape tensor. With \re{vgy} and \re{vhg}, it is in fact straightforward
to find
 \eqn{vcf}{
  \dot{\biTTT\VVA} &= \VVL \VVA + \VVA \VVL^+ \,.
 }
Comparing this with \re{vfw} shows that what we have at hand is actually
the frame free, reference time free, reference configuration free
generalization of the left Cauchy-Green deformation tensor. Its material
version is, therefore, the reformed version of the right Cauchy-Green
tensor.

Based on the definition of \mm{ \VVA \,,} one can show that
 \eqn{vds}{  
  \det \s{{\VVA}_{\bit\Vt}} = \s{\det {\VVA}_{\bit\Vt}} =
  \f{ \s{\p6{\VVh}} }{ \s{\p6{\VVg_{\bit\Vt}}} } =
  \f{ \s{\ppp16{\Mater{\VVh}_{\bit\Vt}}} }{ \s{\p6{\Mater{\VVg}}} } \,,
 }  
where \mm{\p6{\s{\bitt\cdot\bitt}}} denotes the volume form that a
Riemann metric defines. This quantity is, therefore, the ratio of the
current volume and the relaxed volume of a small amount of material. 

For small deformedness, which can already be formulated as \mm{
\VVA_{\bit\Vt} \approx \VVI_{\biTT\VVS} \,,} \ie their difference being much
smaller than \mm{ \VVI_{\biTT\VVS} } in tensorial norm, the rate
equation \re{vcf} becomes, in leading order of
\mm{ \VVA_{\bit\Vt} - \VVI_{\biTT\VVS} \,,}
 \eqn{vck}{ 
  ( {\VVA}_{\bit\Vt} - \VVI_{\biTT\VVS} \Dot{)}
  \approx 2 \VVL\symm \,.
 } 
Comparing this with \re{vhd} shows that we have found the link
 \eqn{vhk}{ 
  \fff{1}{2} \pp1{ {\VVA}_{\bit\Vt} - \VVI_{\biTT\VVS} } \approx
  \VVE^{\bit\hbox{\scriptsize in.\,{Cauchy}}}
 \qquad
   \pp1{ \VVA_{\bit\Vt} - \VVI_{\biTT\VVS} \ll \VVI_{\biTT\VVS} }
 } 
with the Cauchy deformedness \p1{\ie from any given small initial
deformedness, the two evolve the same way in leading order}.

\ssect{vhj}{Deformedness}

For deformedness itself, we are interested in such a function of
\mm{\VVA} that, for \mm{ \VVA_{\bit\Vt} \approx \VVI_{\biTT\VVS} \,,} is
approximately \mm{\fff{1}{2} \pp1{ {\VVA}_{\bit\Vt} - \VVI_{\biTT\VVS} } \,.}
This is actually a large amount of freedom but we already have a
geometrically distinguished desire as well: that the trace of the
deformedness tensor should express the volumetric aspect of elastic
expandedness and its deviatoric part the constant--volume
deformednesses. In the light of Sect.~\ref{vbi} and formula \re{vds}, it
is not surprising to find that this can be accomplished by the Hencky
type choice:
 \eqn{abf}{
  \VVD := \ln \s{\VVA} = \ff{1}{2} \ln \VVA \,,
  \qquad  \Longrightarrow  \qquad
  \tr \bi \VVD = \ln \s{\det \VVA} \,.
 }
As such, \mm{\VVD} also satisfies the wish raised in Sect.~\ref{vbi}
that our deformedness quantity should diverge both for infinite
compression and for infinite extension.

In addition, it is easy to see that
 \eqn{vhl}{
  \VVD \approx \fff{1}{2} \pp1{ {\VVA} - \VVI_{\biTT\VVS} } \approx
  \VVE^{\bit\hbox{\scriptsize in.\,{Cauchy}}}
 \qquad
   \pp1{\hbox{whenever } \VVD \ll \VVI_{\biTT\VVS} } \,.
 }

Although the present work concentrates on the kinematic issues, some
arguments can already be added on the advantage of the Hencky
deformedness in constitutive relations. On the theoretical side, one can
show \p8{the details of the calculation omitted here} that, if
the Cauchy stress \mm{\FFsig} can be derived from an
elastic potential \mm{U\p1{\VVD}} of an isotropic solid as
 \eqn{vhm}{
  \FFsig = \rho \bit \f{\d U}{\d \VVD}
 }
\p1{\mm{\rho} denoting density} then its mechanical power,
\mm{\txt02{tr} \FFsig \biTT \VVL\,,} is found to be
 \eqn{vhn}{
  \txt02{tr} \FFsig \biTT \VVL = \rho \bit \dot U \,.
 }
This is exactly the expression that is the natural one in connection
with a balance equation for energy. In case of any other definition of
deformedness, different from the Hencky one \re{abf}, \re{vhm} should be
replaced by a more complicated expression to retain \re{vhn}. Hence,
this simple example suggests that the Hencky deformedness behaves
distinguishedly from energetic and, consequently, thermodynamic point of
view.

In parallel, it has already been mentioned in Sect.~\ref{vbh} that the
Hencky choice is favoured by experiments in the linear regime of elastic
stress-deformedness relation, as well as for the role of the variable of
nonlinear elastic constitutive relations. Naturally, these aspects
require extensive further investigation to collect a satisfactory amount
of knowledge on the subject.

\ssect{vhi}{The compatibility condition}

We have seen, in Sect.~\ref{var}, that the relaxed metric must be
Ricci-flat. This imposes a restriction on \mm{\VVA \,.} Working with the
spacelike spacetime tensorial version \mm{\VVg_\Vt}, its Ricci tensor
can be expressed in a coordinate free version of the usual coordinate
formula \cite{ONe83b} since the manifold, \mm{\Vt}, on which
\mm{\VVg_\Vt} is given is a Euclidean affine space:
 \eqn{vcs}{
  \VVR^{\hbox{\scriptsize{Ricci}}}_\Vt = \nabla \cdot \VVGam_\Vt - \p1{
  \tr_{1,3} \bit \VVGam_\Vt} \diad \nabla + \p1{ \tr_{1,3} \bit
  \VVGam_\Vt} \bit \VVGam_\Vt - \tr_{1,3} \bit \p1{\VVGam_\Vt
  \VVGam_\Vt} \,,
 }
where the Christoffel tensor is
 \eqn{vct}{
  \VVGam_\Vt = \fff{1}{2} \bitt \VVg_{\bit\Vt^{}}\inv \pp2{
  \VVg_{\bit\Vt^{}} \diad \nabla + \p1{\VVg_{\bit\Vt^{}}
  \diad \nabla}\transs{2,3} - \nabla \diad \VVg_{\bit\Vt^{}} } .
 }
As \mmm{ \VVg_{\bit_\Vt^{}} =\VVh \bit {\VVA}_{\bit\Vt}\inv }
\p2{cf. \re{vcd}}, the requirement \mmm{
\VVR^{\hbox{\scriptsize{Ricci}}}_\Vt = {\bf 0}} can be reformulated as a
condition on \mm{ {\VVA}_{\bit\Vt} }. Omitting the rather lengthy
details of the calculation \cite{xAct}, one can obtain
 \eqn{vcv}{
  \tr_{1,5;3,4} \pp2{ \VVA\inv \diad \VVA\inv \pp1{ \VVA \diad
  \nabla } \pp1{ \VVA \diad \nabla } }
  + 2 \VVh \bit \VVA\inv \bitt \tr_{1,2} \pp2{ \VVA \VVh\inv \pp1{
  \nabla \diad \nabla \diad \VVA } } \VVA\inv &+
 \nnl{vcy}
  + 2 \VVh \bitt \tr_{2,3} \pp2{ \VVA\inv \diad \pp1{\nabla \cdot \VVA}
  \VVh\inv \pp1{ \nabla \diad \VVA } } \VVA\inv
  - 2 \VVh \bit \VVA\inv \bitt \tr_{2,4} \pp2{ \pp1{ \VVA \diad \nabla
  \diad \nabla } \VVh\inv } &+
 \nnl{vcw}
  + 2 \bit \tr_{1,4;3,5} \pp2{ \VVA\inv \diad \VVA\inv \pp1{
  \VVA \diad \nabla } \pp1{ \VVA \diad \nabla } }
  + \tr_{1,2;3,5} \pp2{ \VVA\inv \biT \pp1{ \VVA \diad \nabla }
  \pp1{ \VVA \diad \nabla } \diad \VVA\inv } &-
 \nnl{vcx}
  - 2 \VVh \VVA\inv \bitt \tr_{2,5;3,6} \pp2{ \pp1{ \VVA \diad \nabla }
  \VVA \VVh\inv \pp1{ \nabla \diad \VVA } \diad \VVA\inv } \VVA\inv &+
 \nl{vcz}
  + 2 \bit \tr_{2,4} \pp2{ \pp1{ \nabla \diad \VVA } \VVA\inv
  \pp1{ \VVA \diad \nabla } } \VVA\inv
  - 3 \bit \tr_{2,6;3,5} \pp2{ \pp1{ \nabla \diad \VVA } \VVA\inv
  \pp1{ \VVA \diad \nabla } \diad \VVA\inv } &-
 \nnl{vda}
  - 2 \pp2{ \nabla \diad \pp1{ \nabla \cdot \VVA } } \VVA\inv
  - \VVh \bitt \tr_{3,4;2,5} \pp2{ \VVA\inv \diad \VVA\inv \pp1{ \VVA
  \diad \nabla } \VVA \VVh\inv \pp1{ \nabla \diad \VVA } } \VVA\inv &+
 \nnln{vdb}
  + 2 \bit \tr_{1,2} \pp2{ \VVA\inv \pp1{\VVA \diad \nabla \diad \nabla} }
  - 2 \VVh \bit \VVA\inv \bitt \tr_{2,4;3,5} \pp2{ \pp1{ \VVA \diad
  \nabla } \diad \VVh\inv \pp1{ \nabla \diad \VVA } } \VVA\inv
  &= \bf 0 \,.
 }
This complicated nonlinear equation is to be satisfied by \mm{\VVA \,.}
\quot{Fortunately}, it is enough to satisfy it at a given time and then
the time evolution equation will ensure that at later times it still
holds true.

In the small deformedness regime, the leading order term of this condition
turns out to be extremely simple:
 \eqn{vdh}{
  \nabla \times \VVD
  \times \nabla + \hbox{\p1{higher order terms}} = {\bf 0} \,.
 }
In fact, this is what we would expect even without calculation, as being
the spacetime version of the Saint-Venant compatibility condition
\re{vgc}. Formula \re{vcx} is, hence, the compatibility condition for
finite deformedness. The derivation of \re{vdh} from
\re{vcx} proceeds as follows.
In  \re{vcx}, there are two types of terms: either containing a second
derivative of \m{\VVA} or containing the product of two first
derivatives of \m{\VVA}. Assuming that \mm{ \VVA
- \VVI_{\biTT\VVS} } is small  even in the sense that the product of two first
derivatives is much smaller than a second derivative, we can keep in
\re{vcx} only the terms with a second derivative:
 \eqn{vdd}{
  2 \VVh \bit \VVA\inv \bitt \tr_{1,2} \pp2{ \VVA \VVh\inv \pp1{
  \nabla \diad \nabla \diad \VVA } } \VVA\inv
  - 2 \VVh \bit \VVA\inv \bitt \tr_{2,4} \pp2{ \pp1{ \VVA \diad \nabla
  \diad \nabla } \VVh\inv } &-
 \nln{vde}
  - 2 \pp2{ \nabla \diad \pp1{ \nabla \cdot \VVA } } \VVA\inv
  + 2 \bit \tr_{1,2} \pp2{ \VVA\inv \pp1{\VVA \diad \nabla \diad \nabla} }
  &\approx {\bf 0} \,.
 }
Inserting \mm{ \VVA \approx \VVI_{\biTT\VVS} } directly as well yields
 \eqn{vdg}{
  2 \VVh \bitt \tr_{1,2} \pp2{ \VVh\inv \pp1{
  \nabla \diad \nabla \diad \VVA } }
  - 2 \VVh \bitt \tr_{2,4} \pp2{ \pp1{ \VVA \diad \nabla
  \diad \nabla } \VVh\inv } &-
 \nln{vdf}
  - 2 \pp2{ \nabla \diad \pp1{ \nabla \cdot \VVA } }
  + 2 \bit \tr_{1,2} \pp2{ \pp1{\VVA \diad \nabla \diad \nabla} }
  &\approx {\bf 0} \,.
 }
This formula is already linear in \mm{ \VVA }. Substituting \re{vhl}
into it, and performing simple technical steps, we arrive at \re{vdh}.

For technical purposes, it would be advantageous to be able to
characterize those \mm{\VVA} which satisfy the compatibility condition
in some alternative way, in a way that is simpler than the condition
itself. An idea to do this is as follows. The relaxed metric
\mm{\Mater\VVg} is flat so, for any \mm{\Vt}, it can be brought into
isometry with \mm{\VVh}, the metric of any \mm{\Vt}. Let us choose such
an isometry
 \eqn{vho}{
  \hat{\Vr}_{\Vt} : \VVVVC \to \Vt
 \qquad
  \p1{\forall \Vt} \,.
 }
It looks like a motion of the continuum in spacetime but its is not the
true motion but only an auxiliary \quot{pseudo-motion}. It can be chosen
fairly arbitrarily: actually, it is arbitrary up to the isometries of an
Euclidean affine space, which are the translations and rotations. In
formula, any other such isometry \mm{ \hat{\Vr}'_{\Vt}} is related to
\mm{ \hat{\Vr}_{\Vt}} as
 \eqn{vfa}{
  \hat{\Vr}'_\Vt = \VVR_\Vt \pp1{ \hat{\Vr}_\Vt - \Vo_\Vt } + \Vo_\Vt
 }
with an arbitrary---world line function-like---function
\mm{\Vo_\Vt} \p1{carrying the
translation freedom} and an arbitrary \mm{\VVh}-orthogonal tensor valued
function \mm{\VVR_\Vt}, which embodies the rotation degree of freedom,
and rotates around \mm{\Vo_\Vt \,.} For the Jacobian of the auxiliary
pseudo-motion,
 \eqn{vhp}{
  \hat{\biTT\VVJ}_{\biTT\Vt} := \hat{\Vr}_{\Vt} \diad \nabla \,,
 }
a part of the freedom disappears and only the rotation remains:
 \eqn{vhq}{
  \hat{\biTT\VVJ}_{\biTT\Vt}' = \VVR_\Vt \bitttt
  \hat{\biTT\VVJ}_{\biTT\Vt} \,.
 }
All in all, this auxiliary pseudo-motion is introduced not for any
principal role but as a technical aid.

The property that it is an isometry between \mm{\VVVVC} with
\mm{\Mater\VVg} and \mm{\Vt} with \mm{\VVh} provides the relationship
 \eqn{vdj}{ 
  \Mater{\VVg} = \hat{\biTT\VVJ}_{\biTT\Vt}^* \bi \VVh \bitttt
  \hat{\biTT\VVJ}_{\biTT\Vt}^{}
 }
between the two metrics, and, similarly,
substituting this into \re{vcd} using \re{ver} yields
 \eqn{vhr}{
  \VVA_\Vt = \VVJ_{\biTT\Vt}^{}\bitttt \hat{\biTT\VVJ}_{\biTT\Vt}^{}
  {}\inv \bit \VVh\inv \ppp11{\hat{\biTT\VVJ}_{\biTT\Vt}{}\inv}^*
  \bi \VVJ_{\biTT\Vt}^* \bit \VVh \,.
 }
Let us observe that
 \eqn{vhs}{
  \VVJ_{\biTT\Vt}^{}\bitttt \hat{\biTT\VVJ}_{\biTT\Vt}^{} {}\inv =
  \ppp11{\Vr_\Vt \diad \Mater{\nabla}} \ppp11{\hat{\Vr}_\Vt \diad
  \Mater{\nabla}}\inv = \p0{ \ppp11{\Vr_\Vt \circ
  \hat{\Vr}_\Vt\bit\inv} \diad {\nabla}} = {\hat{\VVq}}_\Vt \diad \nabla
 }
with
 \eqn{vdm}{
  {\hat{\VVq}}_\Vt := \Vr_\Vt \circ \hat{\Vr}_\Vt\bit\inv - \VI_\Vt \,,
 }
a spacelike four-vector field on \mm{\Vt}, where \mm{\VI_\Vt} is the
identity affine map of the affine space \mm{\Vt \,.} In terms of this
vector field,
 \eqn{vdo}{
  {\VVA}_\Vt &=
  \pp1{ {\hat{\VVq}}_\Vt \diad \nabla} \pp1{ {\hat{\VVq}}_\Vt
  \diad \nabla}^+  \,.
 }
Therefore, any \mm{\VVA_\Vt} allowed by the compatibility condition can
be given in terms of a spacelike vector field \mm{{\hat{\VVq}}_\Vt} on \mm{\Vt}.
This \mm{{\hat{\VVq}}_\Vt} is not unique but carries the arbitrarinesses that
hold for \mm{\hat{\Vr}_\Vt \,.} A \mm{{\hat{\VVq}}_\Vt} can be called a vector
potential of elastic shape.

We are actually more familiar with this \mm{{\hat{\VVq}}_\Vt} than it seems at
first sight. Indeed, when
 \eqn{vht1}{
  {\hat{\VVq}}_\Vt \diad \nabla \approx \VVI_{\biTT\VVS}
 }
\p1{in the sense as for the above approximations}
then
 \eqn{vht2}{
  \VVA_\Vt \approx \VVI_{\biTT\VVS}
 }
so we are in the small deformedness regime, and, keeping one more order,
 \eqn{vht3}{
  \VVA_\Vt \approx \VVI_{\biTT\VVS} + 2 \pp1{ {\hat{\VVq}}_\Vt \diad \nabla -
  \VVI_{\biTT\VVS}}\symm ,
 \qquad  \hbox{or}  \qquad
  \VVD_\Vt \approx \pp1{ {\hat{\VVq}}_\Vt \diad \nabla - \VVI_{\biTT\VVS}}\symm .
 }
Let us now take a look at \re{vhv} to realize that \mm{{\hat{\VVq}}_\Vt} is
nothing but the finite deformedness generalization and frame and
etc.~free generalization of the
Cauchy potential.

\sect{vhw}{Plastic kinematics}

As analyzed in Sect.~\ref{vbl}, plasticity is a situation when the
relaxed structure of a solid becomes time dependent.
Accordingly, plastic kinematics must be described by such a quantity
that measures the rate of change of this time dependence.

On the material manifold, this is expressed by
\mm{\dot{\Mater\VVg}_\Vt}---or, equivalently, by
\mm{\big({\Mater\VVg}_\Vt\inv\Dot{\big)} \,.} On spacetime, the
corresponding quantity, \mm{ \VVJ_{\biTT\Vt}^{} \bitt
\big({\Mater\VVg}_\Vt\inv\Dot{\big)} \bit \VVJ_{\biTT\Vt}^* \,,}
provides the additional term that appears when \re{vcf} is generalized
to time dependent \mm{\Mater\VVg}:
 \eqn{vik}{
  \dot{\biTTT\VVA}_\Vt^{} &= \big( \VVJ_{\biTT\Vt}^{} \bitttt
  {\Mater\VVg}_\Vt\inv \VVJ_{\biTT\Vt}^* \VVh \Dot{\big)} = \VVL_\Vt^{}
  \VVA_\Vt^{} + \VVA_\Vt^{} \VVL^+_\Vt - \VVW_{\biTT\Vt}^{}
 }
with
 \eqn{vil}{
  \VVW_{\biTT\Vt}^{} := - \VVJ_{\biTT\Vt}^{} \bitt
  \big({\Mater\VVg}_\Vt\inv\Dot{\big)} \bit \VVJ_{\biTT\Vt}^* \VVh =
  \VVJ_{\biTT\Vt}^{} \bittt {\Mater\VVg}_\Vt\inv \bit
  \dot{\Mater\VVg}{}_\Vt^{} \bit {\Mater\VVg}_\Vt\inv \VVJ_{\biTT\Vt}^*
  \VVh \,.
 }
\mm{\VVW} is a state variable that expresses plastic change rate, and
can be named the {\it metric change tensor}.
Therefore, in a continuum theory involving plastic phenomena, the
complete set of dynamical and constitutive equations must contain an
equation that determines the current value of \mm{\VVW} as a function of
other quantities of the continuum. For example, \mm{\VVW} may be given
as a function of the stress tensor. This function may be nonzero only
when a stress value, \eg the shear stress, exceeds a critical value.

The form of this \mm{\VVW}-determining function might limited by the
requirement that \mm{\VVW} must be such that the relaxed metric
structure \mm{{\Mater\VVg}_\Vt} remains Ricci-flat at any time \m{\Vt}.
This would correspond to the physical expectation that a sudden
unloading should stop plastic changes, it should stop the change of the
relaxed metric and---in case of zero volume forces---bring the material
to an elastically nondeformed state, bring the current metric to the
stressless one.

Bearing in mind the complicated form of the compatibility condition
\re{vcx}, to ensure that the time dependent relaxed metric stays
Ricci-flat is a rather strong restriction on \mm{\VVW} mathematically. On
the other side, physically it might be valuable since many candidates for
the \mm{\VVW}-determining function that seem reasonable for small
deformations may fail to admit a satisfying finite deformation
generalization. The strong requirement---if turns out to be physically
justified---helps to find physically admissible models by allowing only a
rather limited range of choices.

The two widely used formulas for plastic kinematics, \mmm{
\DDH^{\txt00{total}} = \DDH^{\txt00{elast}} \bitt \DDH^{\txt00{plast}} }
and \mmm{ \DDE^{\txt00{total}} = \DDE^{\txt00{elast}} +
\DDE^{\txt00{plast}} } \p2{appeared in Sect.~\ref{vac} under \re{vft}
and \re{vfs}, respectively, and already discussed in
Sects.~\ref{vbl}--\ref{vbm}} can be viewed from the kinematic picture
presented here as follows.

Let us rewrite \re{vhs} as
 \eqn{vib}{
  \VVJ = \p0{\Vr \diad \Mater{\nabla}} =
  \pp1{{\hat{\VVq}} \diad \nabla} \ppp11{\hat{\Vr} \diad \Mater{\nabla}}
  = \pp1{{\hat{\VVq}} \diad \nabla} \bitt \hat{\biTT\VVJ}
 }
The Jacobian tensor \mm{\VVJ} has already been found to be the spacetime
compatible generalization of the deformation gradient \mm{\DDH = \DDH^{\txt00{total}}
\,.} Furthermore, the pseudo-motion \mm{\hat{\Vr}} is a mapping of the
relaxed metric structure into spacetime at any instant so it can be
\quot{interpreted} \p1{pseudo-interpreted} as a \quot{motion of relaxed
positions}. In this sense, \mmm{\hat{\Vr} \diad \Mater{\nabla} =
\hat{\biTT\VVJ}} is the incarnation of \mmm{\DDH^{\txt00{plast}}\,.}
Finally, \mmm{{\hat{\VVq}} \diad \nabla} has been seen to be related to the
elastic shape tensor \mm{\VVA} --- compare \re{vdo} with the usual
formula \mmm{ \DDUL^2 = \DDH^{\txt00{elast}} \bitt
\p1{\DDH^{\txt00{elast}}}\transp } of Sect.~\ref{vac} and recall that
\mm{\DDUL^2} is the traditional counterpart of \mm{\VVA \,.} Hence,
\mmm{{\hat{\VVq}} \diad \nabla} embodies \mmm{\DDH^{\txt00{elast}}\,,} and thus
the relationship between \mmm{ \DDH^{\txt00{total}} = \DDH^{\txt00{elast}} \bitt
\DDH^{\txt00{plast}} } and \re{vib} becomes clear.

It is also apparent that the unphysical arbitrarinesses of
\mm{\DDH^{\txt00{elast}}} and \mm{\DDH^{\txt00{plast}}}, pointed out
in Sect.~\ref{vbm}, appear here as the unphysical arbitrarinesses in the
pseudo-motion \mm{\hat{\Vr}} and in the related quantities
\p1{\mm{\hat{\biTT\VVJ}} and \mm{{\hat{\VVq}}}}.

Next, let us investigate \re{vik} in the regime of small \emp{elastic}
deformedness. \p1{Note that it has no meaning to consider \quot{small
plastic deformedness} because there is no plastic
deformedness/deformation/strain quantity as a state variable.}
 In the leading order of small \mmm{ \VVA - \VVI_{\biTT\VVS} \,,} \re{vik}
simplifies to
 \eqn{vie}{
  2 \dot{\VVD}_\Vt^{} &\approx \VVL_\Vt^{} + \VVL^+_\Vt +
  \VVJ_{\biTT\Vt}^{} \bitt \big({\Mater\VVg}_\Vt\inv\Dot{\big)} \bit
  \VVJ_{\biTT\Vt}^* \VVh = \pp1{ \VVv_\Vt^{} \diad \nabla } + \VVh\inv
  \biT \pp1{ \nabla \diad \VVv_\Vt^{} } \VVh + \VVJ_{\biTT\Vt}^{} \bitt
  \big({\Mater\VVg}_\Vt\inv\Dot{\big)} \bit \VVJ_{\biTT\Vt}^* \VVh \,,
 }
which can be rearranged as
 \eqn{vif}{
  \f{1}{2} \pp2{ \ppp11{ \VVv_\Vt^{} \diad \Mater{\nabla} } \VVJ\inv_\Vt
  + \VVh\inv \biT \p1{\VVJ^*_\Vt}\inv \ppp11{ \Mater{\nabla} \diad
  \VVv_\Vt^{} } \VVh }  &\approx  \dot{\VVD}_\Vt^{} -\VVJ_{\biTT\Vt}^{}
  \bitt \big({\Mater\VVg}_\Vt\inv\Dot{\big)} \bit \VVJ_{\biTT\Vt}^* \VVh
  \,.
 }

Then, let \mm{\p2{\Vt_1, \Vt_2}} be a time interval during which
\mm{\VVJ} changes only a little:
 \eqn{vim}{
  \VVJ\p1{\Vt'} \VVJ\p1{\Vt}\inv \approx \VVI_{\biTT\VVS}
  \qquad  \p1{ \forall \Vt, \Vt' \in \p2{\Vt_1, \Vt_2} } \,.
 }
 Then we can replace \mm{\VVJ} with
\mm{\VVJ_{\Vt_1}}, for example. Integrating in time over \mm{\p2{\Vt_1,
\Vt_2}} at a fixed material point, and noting that
 \eqn{vig}{
  \int_{\Vt_1}^{\Vt_2} \VVv =  \int_{\Vt_1}^{\Vt_2} \dot{\VVr} =
  \Vr_{\Vt_2} - \Vr_{\Vt_1} =: \Delta \Vr \,,
 }
we obtain
 \eqn{vii}{
  \f{1}{2} \pp2{ \ppp11{ \Delta \Vr \diad \Mater{\nabla} }
  \VVJ\inv_{\Vt_1} + \VVh\inv \biT \ppp11{\VVJ^*_{\Vt_1}}\inv \ppp11{
  \Mater{\nabla} \diad \Delta \Vr } \VVh }  & \approx  \Delta \VVD -
  \VVJ_{\biTT\Vt_1}^{} \pp2{ \int_{\Vt_1}^{\Vt_2}
  \big({\Mater\VVg}_\Vt\inv\Dot{\big)} }
  \VVJ_{\biTT\Vt_1}^* \bitt \VVh
 \nnl{vin}
  \approx  \Delta \VVD - \VVJ_{\biTT\Vt_1}^{} \pp2{
  {\Mater\VVg}_{\Vt_2}\inv - {\Mater\VVg}_{\Vt_1}\inv }
  \VVJ_{\biTT\Vt_1}^* \bitt \VVh  & \approx  \Delta \VVD +
  \Delta \pp1{ - \VVJ \bit {\Mater\VVg}\inv \VVJ^* \bitt \VVh }
 }
\p1{where, again, \mm{\Delta} means changes between \mm{\Vt_1} and
\mm{\Vt_2}}. On the rightmost hand side, \mm{\Delta \VVD} is the change
of elastic deformedness and the second term expresses purely plastic
changes. Therefore, we have reached a formula that embodies \mmm{ \DDE^{\txt00{total}}
= \DDE^{\txt00{elast}} + \DDE^{\txt00{plast}} \,.}

Now, the formalism in
Sect.~\ref{vac} suggests \mm{\DDE^{\txt00{total}}} be some realization of
 \eqn{vij}{
  \f{1}{2} \ppp22{ \ppp11{ \Delta \DDu \diad \nablaR }
  \,\hbox{\quott{+}}\, \ppp11{ \nablaR \diad \Delta \DDu } }
 }
on the leftmost hand side of \re{vin}. However, it has no meaning that
\mm{\VVJ_{\Vt_1}} is approximately identity because \mmm{\VVJ_{\Vt_1}
\p1{\Vp} : \tang{\Vp}{\VVVVC} \to \VVS } connects different vector
spaces, between which there is no {identity} map. Only taking \m{\Vt_1}
as a reference time \p8{and thus choosing an identification between
\mm{\tang{\Vp}{\VVVVC}} and \mm{\VVS}} enables one to transform
\mm{\VVJ_{\Vt_1}} into the quantities which measure changes with respect
to the reference time. The same step is needed to give a meaning to the
addition in \re{vij}, too, since an \mm{ \VVS \diad \tang{\Vp}{\VVVVC}^*
} tensor and a \mm{ \tang{\Vp}{\VVVVC}^* \diad \VVS } tensor cannot be
added---that's why the + in \re{vij} is put within quotation marks.
Furthermore, \mm{\Delta \DDu} is defined only with respect to a
reference frame. Without reference elements and artificial
identification, \re{vij} is not meaningful and, for a \mm{
\DDE^{\txt00{total}} } quantity, one has to be content with the leftmost
side of \re{vin}.

\sect{via}{Conclusion}

The paper has elaborated and emphasized the following points:
 \begin{enumerate}
 \item
Continuum physics must avoid the use of auxiliary elements like
reference frame, reference time and reference configuration, which cause
that the observed gets mixed with the observer and \textit{unphysical
artefacts} emerge in the description.
 \item
If working solely with material manifold based quantities, one cannot
give account of inertial-noninertial aspects of the motion/flow of the
continuum.
 \item
By \textit{working on spacetime}, one is able to introduce the necessary
kinematic continuum quantities without any auxiliary reference elements,
and the inertial-noninertial aspects are also properly included. The
real contents of the question of material frame indifference is
clarified as the question of \textit{noninertial motion indifference}.
 \item
In the spacetime-based objective approach, the motion of a material
point is given not as time dependent position in a reference frame space
but as a world line in spacetime.
 The spacetime compatible incarnation of the usual deformation gradient
quantity is the gradient of the system of material world lines. This
\textit{world line gradient} is defined for solids, liquids and gases
each.
 \item
All these types of continua also admit a \textit{current metric} on the
material manifold. This metric expresses the momentary distance of any
two material points in spacetime, for any instant.

In addition, solids \p8{but only solids} possess another Riemannian
metric as well: the \textit{stressless} or \textit{relaxed metric}, which
tells the distance of any two material points when the body is in an
unloaded and fully relaxed state.
 It follows that the relaxed metric must have zero Riemann curvature
tensor and zero Ricci tensor. This requirement is the \quot{finite
deformation compatibility condition}.
 \item
Elastic stress is expected to depend on how the current pairwise
distances of material points differ from the relaxed distances.
Correspondingly, the elastic kinematic quantity is a \textit{state
variable} \p1{as opposed to being a two-point function}, which compares
the current metric with the relaxed metric. The mathematically natural
quantity comparing the two metrics, named the \textit{elastic shape
tensor}, proves to be the spacetime generalization of the left
Cauchy-Green deformation tensor.
 \item
The elastic shape tensor evolves in time via a \textit{rate equation}.
The solution of this differential equation can be fixed by an initial
condition, which, in many practical cases, is not the relaxed state.
 \item
For the purpose of the objective version of finite deformation elastic
strain, the \textit{elastic deformedness tensor} is defined from elastic
shape as a spacetime analogue of the left Hencky strain. It is
kinematically distinguished by that its trace expresses volumetric
expansion and its deviatoric part the shear degrees of freedom. Certain
physical results also favour this logarithmic choice.
 \item
\textit{Plasticity} is a phenomenon when \textit{the relaxed metric}
structure of the solid \textit{changes}.
The change rate of the relaxed
metric is to be determined by a plastic constitutive equation.
 \item
No \quott{elastic deformation gradient} and no \quott{plastic
deformation gradient} can be defined objectively. The typically used
additive and multiplicative rules \mmm{ \DDE^{\txt00{total}} =
\DDE^{\txt00{elast}} + \DDE^{\txt00{plast}} } and \mmm{
\DDH^{\txt00{total}} = \DDH^{\txt00{elast}} \bitt \DDH^{\txt00{plast}} }
for \quot{elastic and plastic deformation}, resp. \quot{elastic and
plastic deformation gradient}, fail to possess a spacetime compatible
and free-of-auxiliary-reference-elements meaning.
 \end{enumerate}

\section{Acknowledgement}

The authors thank J. M. Mart\'\i n-Garc\'\i a for his help in using the
{\it xAct} software packages \cite{xAct}. Illuminating discussions with
Csaba Asszonyi and Arkadi Berezovski are gratefully acknowledged. The
work was supported by the grant OTKA K81161.

\appendix

\sect{vgg}{Spacetime without reference frames}

\ssect{vgh}{Quantities with physical dimension \p1{length, time, etc.}}

Units chosen for dimensionful physical quantities \p8{length, time,
mass, etc.} are also auxiliary elements in a description. Therefore, let
us avoid them, too, by utilizing the unit free approach \cite{Mat93b}, where
we introduce separate one dimensional oriented real vector spaces \p1{in
short, measure lines} \mm{\VVVL\,,} \mm{\VVVT\,,} \mm{\VVVM} for
lengths, time intervals and mass values, respectively. A length, for
example, is an element \mm{\VVlen} of \mm{\VVVL}, sums of lengths
\mm{\VVlen_1 + \VVlen_2} and real multiples of lengths \mm{\alpha
\VVlen} also being elements of \mm{\VVVL \,.} It will be useful to
summarize here how the unit free treatment gives account of the
properties of dimensionful quantities.

The product of a length \mm{\VVlen \in \VVVL} and a time \mm{\VVt \in
\VVVT} is their tensorial product \mmm{\VVlen \diad \VVt \in \VVVL \diad
\VVVT} and their quotient is the tensorial quotient \mmm{\VVlen \Per
\VVt \in \VVVL \Per \VVVT\,,} with all the natural and expected
properties, like
 \eqn{vgl}{
  \pp1{ \VVlen_1 + \VVlen_2 } \diad \VVt &=
  \VVlen_1 \diad \VVt + \VVlen_2 \diad \VVt \,,
 &
  \pp1{ \VVlen_1 + \VVlen_2 } \Per \VVt &=
  \VVlen_1 \Per \VVt + \VVlen_2 \Per \VVt \,,
 \nl{vgm}
  \pp1{ \alpha \VVlen } \diad \pp1{ \beta \VVt } &=
  \pp1{ \alpha \beta } \bitt \p0{ \VVlen \diad \VVt } \,,
 &
  \pp1{ \alpha \VVlen } \Per \pp1{ \beta \VVt } &=
  \pp1{ \alpha / \beta } \bitt \p0{ \VVlen \Per \VVt } \,.
 }
The tensorial power of a measure line is defined for any real power and
is also a measure line, \lat{e.g.,} the \m{p}\/th power of \mm{\VVVL} is
\mm{\VVVL^{\VVVpow{p}}}. For \mm{\VVlen \in \VVVL \,,}
\mm{\VVlen^{\VVVpow{p}} \in \VVVL^{\VVVpow{p}} \,,} and properties like
\mm{ \p1{ \p6{\alpha} \VVlen }^{\VVVpow{p}} = \p6{\alpha}^p
\VVlen^{\VVVpow{p}} } are satisfied. For positive integer powers,
tensorial power coincides with what is straightforwardly expected,
\ie \mmm{\VVVL^{\VVVpow{2}} = \VVVL \diad \VVVL \,,}
\mmm{\VVVL^{\VVVpow{3}} = \VVVL \diad \VVVL \diad \VVVL \,,} etc.
Furthermore, \mm{\VVVL^{\VVVpow{0}}} is the vector space of real
numbers, \mm{\VVVR}, and \mm{\VVVL^{\VVVpow{-1}}} turns out to equal
\mm{\VVVR \Per \VVVL \equiv \VVVL^* \,,} where \mm{^*} denotes the dual
vector space.

If \mm{\VVA} is a linear map between vector spaces,
\mmm{ \VVA: \VVU \to \VVV \,,} then \p1{taking again \mm{\VVVL} as an
example} \mm{\VVA} also determines a linear map \mm{\VVU \diad \VVVL \to
\VVV \diad \VVVL}
\p1{if \mm{ \VVA: \VVu \mapsto \VVv } then the corresponding map does
\mmm{ \VVu \diad \VVlen \mapsto \VVv \diad \VVlen } for all \mmm{\VVlen
\in \VVVL}}. It will not cause confusion to denote this corresponding
map also by \mm{\VVA \,.}

Continuity, derivative, integral etc.\ of measure line valued functions,
and of functions of measure line variable, is analogous to that of
\mmm{\VVVR \to \VVVR} functions, with the expected properties. For
example, the derivative of a function \mmm{\VVVT \to \VVVL} is \mmm{
\VVVL \Per \VVVT } valued, \ie has the dimension of velocity.

Units are not needed when formulating physical theories. Nevertheless,
for completeness, it is worth closing this brief summary with
how units are included in this approach. Choosing a unit for length is
choosing a basis vector \mm{\VVlen_0} in the one dimensional vector
space \mm{\VVVL \,,} with which any length \mm{\VVlen} can be given in a
form \mm{\VVlen = \lambda \VVlen_0} and, consequently, can be
characterized by a real number, the uniquely determined \mm{\lambda\,.}
Change to use a new unit means change to a new basis, to a new single
basis vector.

\ssect{vgi}{The structure of nonrelativistic spacetime}

\def\cca{480}  \def\ccb{340}  
\def\ccc{80}   \def\ccd{25}   
\def\cce{80}   \def\ccf{170}  
\def\ccg{2}    \def\cch{3}    
\setadd{\cci}{\ccc}{\cce}     
\setmuldiv{\cco}{\cce}{\cch}{\ccg}    
\setaddadd{\ccj}{\ccd}{\ccf}{\cco}    
\set{\ccz}{25}                
\set{\ccz}{45}                
\setadd{\ccv}{\cci}{\ccz}     
\setadd{\ccw}{\ccv}{\cce}     
\setadd{\ccx}{\ccw}{\ccz}     
\setadd{\ccy}{\ccx}{\cce}     
\setdivadd{\cck}{\cce}{2}{\ccv}                  
\setdiv{\ccl}{\cco}{2}  \addadd{\ccl}{\ccd}{69}  
\setsub{\cka}{\ccl}{11}                          
\setsub{\ccm}{\cck}{15} \setadd{\ccn}{\ccl}{14}  
\setadd{\ccp}{\cck}{-3} \setadd{\ccq}{\ccl}{23}  
\setsub{\ccr}{\cck}{6} \setadd{\ccs}{\ccl}{12} 
\setadd{\cct}{\cck}{4} \setadd{\ccu}{\ccl}{14} 
\def\cda{30}  \def\cdb{0}  
\def\cdc{425}              
\def\cdc{430}              
\setaddadd{\cdd}{\cda}{\cdc}{18}     
\setsub{\cde}{\cdb}{22}              
\setsubsub{\cdf}{\cck}{\cce}{\ccz}   
\setaddadd{\cdg}{\cck}{\ccz}{\cce}   
\setadd{\cdi}{\ccv}{12}  \setadd{\cdj}{\ccd}{40}  
\setsub{\cea}{\ccc}{30}\setadd{\ceb}{\ccd}{145}\set{\cfa}{60}\set{\cfb}{49}
\set{\cem}{\ccc} \setadd{\cen}{\ceb}{17}\set{\cfm}{60}\set{\cfn}{32}
\setadd{\ceg}{\cea}{65}\setadd{\ceh}{\ceb}{32}\set{\cfg}{60}\set{\cfh}{23}
\setadd{\cei}{\ceg}{104}\setadd{\cej}{\ceb}{45}\set{\cfi}{60}\set{\cfj}{-13}
\setadd{\cec}{\ccv}{40}\setadd{\ced}{\ceb}{40}\set{\cfc}{60}\set{\cfd}{-18}
\setadd{\cek}{\cei}{117}\setsub{\cel}{\cej}{11}\set{\cfk}{60}\set{\cfl}{8}
\setadd{\ceo}{\cek}{32}\setadd{\cep}{\cel}{5}\set{\cfo}{60}\set{\cfp}{9}
\setadd{\cee}{\ccx}{105}\setadd{\cef}{\ced}{25}\set{\cfe}{60}\set{\cff}{29}
\setsub{\cga}{\ccc}{35}\setadd{\cgb}{\ccd}{70}\set{\cha}{60}\set{\chb}{69}
\set{\cgm}{\ccc}\setadd{\cgn}{\cgb}{31}\set{\chm}{60}\set{\chn}{35}
\setadd{\cgg}{\cga}{65}\setadd{\cgh}{\cgb}{45}\set{\chg}{60}\set{\chh}{20}
\set{\cgi}{\ccv}\setadd{\cgj}{\cgb}{48}\set{\chi}{60}\set{\chj}{-13}
\setadd{\cgc}{\ccv}{38}\setadd{\cgd}{\cgb}{40}\set{\chc}{60}\set{\chd}{-20}
\set{\cgk}{\ccx}\setsub{\cgl}{\cgj}{15}\set{\chk}{60}\set{\chl}{8}
\setadd{\cgo}{\cgk}{32}\setadd{\cgp}{\cgl}{5}\set{\cho}{60}\set{\chp}{18}
\setadd{\cge}{\ccx}{105}\setadd{\cgf}{\cgd}{35}\set{\che}{60}\set{\chf}{39}
\setsub{\cia}{\ccc}{20}\setadd{\cib}{\ccd}{123}\set{\cja}{60}\set{\cjb}{45}
\set{\cim}{\ccc}\setadd{\cin}{\cib}{14}\set{\cjm}{60}\set{\cjn}{35}
\setadd{\cig}{\cia}{65}\setadd{\cih}{\cib}{36}\set{\cjg}{60}\set{\cjh}{15}
\set{\cii}{\ccv}\setadd{\cij}{\cib}{43}\set{\cji}{60}\set{\cjj}{-13}
\setadd{\cic}{\ccv}{50}\setadd{\cid}{\cib}{30}\set{\cjc}{60}\set{\cjd}{-25}
\set{\cik}{\ccx}\setsub{\cil}{\cij}{21}\set{\cjk}{60}\set{\cjl}{8}
\setadd{\cio}{\cik}{48}\setadd{\cip}{\cil}{7}\set{\cjo}{60}\set{\cjp}{18}
\setadd{\cie}{\ccx}{115}\setadd{\cif}{\cid}{24}\set{\cje}{60}\set{\cjf}{27}

\unitlength=.35pt

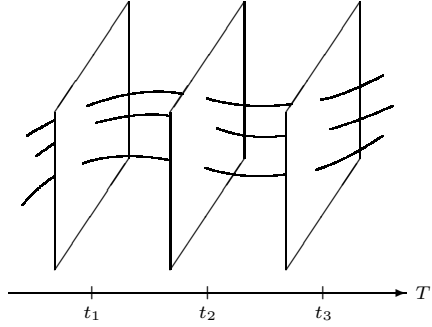
\begin{figure}
\begin{center}
\begin{picture}(\cca,\ccb)(0,0)
\put(\ccc,\ccd){\line(\ccg,\cch){\cce}}
\put(\ccc,\ccd){\line(0,1){\ccf}}
\put(\cci,\ccj){\line(0,-1){\ccf}}
\put(\cci,\ccj){\line(-\ccg,-\cch){\cce}}
\put(\ccv,\ccd){\line(\ccg,\cch){\cce}}
\put(\ccv,\ccd){\line(0,1){\ccf}}
\put(\ccw,\ccj){\line(0,-1){\ccf}}
\put(\ccw,\ccj){\line(-\ccg,-\cch){\cce}}
\put(\ccx,\ccd){\line(\ccg,\cch){\cce}}
\put(\ccx,\ccd){\line(0,1){\ccf}}
\put(\ccy,\ccj){\line(0,-1){\ccf}}
\put(\ccy,\ccj){\line(-\ccg,-\cch){\cce}}
\put(\cda,\cdb){\vector(1,0){\cdc}}
\put(\cdd,\cdb){\nb{\scriptsize$\VT$}}
\put(\cdf,\cde){\nb{\scriptsize$\bit\Vt_1$}}
\put(\cck,\cde){\nb{\scriptsize$\bit\Vt_2$}}
\put(\cdg,\cde){\nb{\scriptsize$\bit\Vt_3$}}
\put(\cck,\cdb){\nb{\tiny\boldmath{$|$}}}
\put(\cdf,\cdb){\nb{\tiny\boldmath{$|$}}}
\put(\cdg,\cdb){\nb{\tiny\boldmath{$|$}}}
\beztan{\cea}{\ceb}{\cfa}{\cfb}{\cem}{\cen}{\cfm}{\cfn}
\beztan{\ceg}{\ceh}{\cfg}{\cfh}{\cei}{\cej}{\cfi}{\cfj}
\beztan{\cec}{\ced}{\cfc}{\cfd}{\cek}{\cel}{\cfk}{\cfl}
\beztan{\ceo}{\cep}{\cfo}{\cfp}{\cee}{\cef}{\cfe}{\cff}
\beztan{\cga}{\cgb}{\cha}{\chb}{\cgm}{\cgn}{\chm}{\chn}
\beztan{\cgg}{\cgh}{\chg}{\chh}{\cgi}{\cgj}{\chi}{\chj}
\beztan{\cgc}{\cgd}{\chc}{\chd}{\cgk}{\cgl}{\chk}{\chl}
\beztan{\cgo}{\cgp}{\cho}{\chp}{\cge}{\cgf}{\che}{\chf}
\beztan{\cia}{\cib}{\cja}{\cjb}{\cim}{\cin}{\cjm}{\cjn}
\beztan{\cig}{\cih}{\cjg}{\cjh}{\cii}{\cij}{\cji}{\cjj}
\beztan{\cic}{\cid}{\cjc}{\cjd}{\cik}{\cil}{\cjk}{\cjl}
\beztan{\cio}{\cip}{\cjo}{\cjp}{\cie}{\cif}{\cje}{\cjf}
\end{picture}%
\end{center}
\y1ex
\caption{The structure of nonrelativistic spacetime}
\label{vgn}
\end{figure}

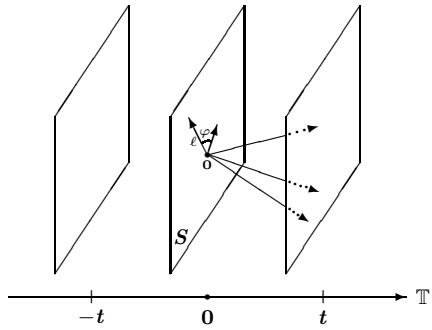
\begin{figure}
\begin{center}
\begin{picture}(\cca,\ccb)(0,0)
\put(\ccc,\ccd){\line(\ccg,\cch){\cce}}
\put(\ccc,\ccd){\line(0,1){\ccf}}
\put(\cci,\ccj){\line(0,-1){\ccf}}
\put(\cci,\ccj){\line(-\ccg,-\cch){\cce}}
\put(\cck,\ccl){\circle*{6}}
\put(\cck,\ccl){\vector(-1,2){20}}
\put(\cck,\ccl){\vector(1,3){11}}

\put(\cck,\ccl){\line(4,1){85}}
\put(365,184){\vector(4,1){0}}
\multiput(334,176.25)(8,2){3}{\circle*{1}}

\put(\cck,\ccl){\line(3,-1){85}}
\put(366,113.3){\vector(3,-1){0}}
\multiput(334,124.4)(6,-2){4}{\circle*{1}}

\put(\cck,\ccl){\line(3,-2){85}}
\put(355,80.67){\vector(3,-2){0}}
\multiput(334,94.6)(4.5,-3){3}{\circle*{1}}

\put(\cck,\cka){\nb{$\zS \tens0$}}
\put(\ccm,\ccn){\nb{$\zS \ell$}}
\put(\ccp,\ccq){\nb{$\bittt\!{\zS \varphi}$}}
\beztan{\ccr}{\ccs}{5}{4}{\cct}{\ccu}{2}{-3}
\put(\ccv,\ccd){\line(\ccg,\cch){\cce}}
\put(\ccv,\ccd){\line(0,1){\ccf}}
\put(\ccw,\ccj){\line(0,-1){\ccf}}
\put(\ccw,\ccj){\line(-\ccg,-\cch){\cce}}
\put(\ccx,\ccd){\line(\ccg,\cch){\cce}}
\put(\ccx,\ccd){\line(0,1){\ccf}}
\put(\ccy,\ccj){\line(0,-1){\ccf}}
\put(\ccy,\ccj){\line(-\ccg,-\cch){\cce}}
\put(\cda,\cdb){\vector(1,0){\cdc}}
\put(\cdd,\cdb){\nb{\footnotesize$\VVVT$}}
\put(\cck,\cdb){\circle*{6}}
\put(\cdf,\cde){\nb{\footnotesize$-\VVt$}}
\put(\cck,\cde){\nb{\footnotesize$\vect0$}}
\put(\cdg,\cde){\nb{\footnotesize$\VVt$}}
\put(\cdf,\cdb){\nb{\tiny\boldmath{$|$}}}
\put(\cdg,\cdb){\nb{\tiny\boldmath{$|$}}}

\put(\cdi,\cdj){\nb{\footnotesize$\VVS$}}

\end{picture}
\ph{--}
\end{center}
\y1ex
\caption{The structure of nonrelativistic spacetime vectors}
\label{vgo}
\end{figure}

Our notations for the reference frame free treatment of nonrelativistic
spacetime \cite{Mat93b} will be as follows. See Figures~\ref{vgn} and
\ref{vgo} for corresponding illustrations.

The set of spacetime points, \mm{\VM}, is a four dimensional real
oriented affine space over \mm{\VVM}, the set of spacetime vectors
\p1{also called four-vectors}. To any two spacetime points \mm{\Fxa,
\Fxb \in \VM } belongs a spacetime vector \mm{\FFx \in \VVM}, in
notation, \mmm{\Fxb - \Fxa = \FFx \,,} \mmm{\Fxb = \Fxa + \FFx \,.}
Absolute time structure of spacetime vectors is given by a nondegenerate
linear map \mm{\VVtau: \VVM \to \VVVT \,,} assigning a time interval to
any spacetime vector. On Fig.~\ref{vgo}, the \quot{slices} depict
equal-\mm{\VVtau} surfaces. The elements of the three dimensional null
space of \mm{\VVtau\,,}
 \eqn{vgp}{
  \VVS := \txt02{Ker} \VVtau =
  \{ \FFx \in \VVM \bitt : \bitt \VVtau \FFx = \vect0 \}
 }
are named spacelike vectors {since their \quot{timelike aspect}, the
value of \mm{\VVtau\,,} is zero}. They are also called three-vectors.
\mm{\VVS} plays two important roles. First, the equivalence relation
 \eqn{vgq}{
  \Fxa \sim \Fxb \makebox[4em]{$\Longleftrightarrow$}
  \Fxb - \Fxa \in \VVS
 }
defines the time instants of spacetime as the equivalence classes in
\mm{\VVM}---the \quot{slices} in Fig.~\ref{vgn}. The set of time
instants is a one dimensional real oriented affine space \mm{\VT \,.}
The time interval evaluation of spacetime vectors, \mm{\VVtau}, induces
a time instant evaluation of spacetime points, which is an affine
surjection \mm{\Vtau : \VM \to \VT \,.} Therefore, expressing in terms
of \mm{\Vtau\,,} the slices in Fig.~\ref{vgn} are the
equal-\mm{\Vtau}-surfaces. To emphasize again: an instant is a
collection of simultaneous spacetime points.

The other important aspect of \mm{\VVS} is regarding the spatial
structure on nonrelativistic spacetime, which is a Euclidean scalar
product, \ie a positive definite symmetric bilinear map \mmm{\VVh : \VVS
\times \VVS \to \VVVL^{\VVVpow{2}} \,.} Note that, nonrelativistically,
only a three dimensional subset of spacetime vectors possesses a
\p8{Euclidean} metric structure, as opposed to the four dimensional
\p8{pseudo-Euclidean} metric structure of special relativistic spacetime
vectors. Lengths and angles \p1{see Fig.~\ref{vgo}} have meaning only
within \mm{\VVS\,.}

Summarizing, nonrelativistic spacetime is a quintuplet \mm{\p1{\VM,
\VVVT, \VVtau, \VVVL, \VVh}}, with the explanations above.

After the defining properties, let us also consider those consequences
and physically important notions which will be utilized for the 
forthcoming discussion of continuum kinematics.

The dual vector space of \mm{\VVM\,,} \mm{\VVM^*\,,} consists of the
linear \mm{ \VVM \to \VVVR } maps. Unlike special relativistically, in
the nonrelativistic case four-covectors cannot be identified with
four-vectors. \p1{See, \eg \cite{Mat93b} for tensorial identifications
and other various notions and notations related to tensors that are
utilized hereafter.} Let \mmm{\VVeta : \VVM^* \to \VVS^*} denote the
linear map that restricts any four-covector to the subspace of spacelike
vectors:
 \eqn{vgr}{
  \VVeta : \VVk \mapsto \left.\VVk\yph{|^!}\right|^{}_{\VVS}
 }
Its dual transpose, \mm{ \VVeta^*: \VVS
\to \VVM } is nothing but
\mm{\left.\VVI\biTT_{\VVM}{\yph{|^!}}\right|^{}_{\VVS} \,,} the
restriction of the identity map of \mm{\VVM } on spacelike vectors.

The motion \p8{or, so to say, the existence} of a pointlike object in
spacetime is described by a world line, a curve with futurelike tangent
vectors, where futurelike means vectors that have positive \mm{\VVtau}
value. \p1{Fig.~\ref{vgn} depicts three world lines.} A world line
function is a world line with its natural parametrization by time:
 \eqn{vgs}{
  \Vr : \VT \to \VM \,,
 \qquad
  \Vtau \p1{ \Vr_{\Vt} } = \Vt  \quad \p1{\forall \Vt \in \VT}
 }
\p1{let's keep writing the time variable into subscript}, fulfilling the
property that its derivative, the four-velocity at any time \mm{\Vt},
satisfies
 \eqn{vgt}{
  \VVtau \dot{\Vr}_\Vt = 1 \,.
 }
For this reason, it is useful to give a notation to the set of all
possible four-velocity values:
 \eqn{vgu}{
  V(1) := \{ \VVv \in \VVM \biT \Per \VVVT \bitt : \bitt \VVtau \VVv = 1 \}
  \,,
 }
which is a three dimensional Euclidean affine space over \mm{\VVS \Per
\VVVT \,.}

The motion \p1{spacetime existence, \quot{fate}} of an extended and
continuous material object is a continuous collection of world lines,
which provides a foliation of spacetime. Then, at any spacetime point,
we have a four-velocity value, which is the derivative there of the
world line function that goes through that spacetime point. Thus a
four-velocity field on spacetime can be read off. The opposite way is
also possible for giving a motion of a continuum, when we start with a
\p8{smooth enough} four-velocity field on spacetime and determine the
maximal integral curves of that four-vector field, which integral curves
will be the world lines of the material body.

Any reference frame or spacetime observer is also such a possible
continuum motion in spacetime. The space points of an observer are the
world lines. Especially, an inertial observer admits world lines that
are parallel straight lines with a single prescribed four-velocity value
as their tangent vector that is constant along each world line and is
the same value for all world lines.

A more general and also important family of observers is the rigid
observers, where the distance of any two space points of the observer
are time independent. Note that an instant is a spacelike three
dimensional affine subspace in spacetime which intersects each world
line, each futurelike curve at exactly one spacetime point, and the
distance of two space points of an observer at a time is the Euclidean
length of the two such intersection points. When this distance is the
same for all times for this given pair of world lines, and if all such
pairwise distances are time independent then the observer is called
rigid.

The space of a rigid observer proves to be a three dimensional Euclidean
affine space. For nonrigid observers, not even a time independent
Riemann metric structure can be found, in general, so we are left with
only a smooth manifold.

Calculus of functions defined on spacetime is similar to that of defined
on \mm{\VVVR^4\,,} though some differences emerge because \mm{\VVVR^4}
is a Euclidean vector space while, on \mm{\VVM}, which lies under
\mm{\VM\,,} the structures \mm{\VVtau} and \mm{\VVh} are more tricky.
The spacetime derivative of a scalar field \mmm{f : \VM \to \VVVR\,,}
denoted by \mm{f \diad \VVVD} \p1{to be systematic, derivatives will
always be written indicating the correct tensorial order} takes an
\mm{\VVM^*} value at each spacetime point. It expresses how fast \mm{f}
changes in a spacetime direction. The restriction \mmm{ \VVeta \p1{f
\diad \VVVD} = \p1{f \diad \VVVD} \VVeta^* \,,} denoted hereafter by \mm{
f \diad \nabla \,,} tells how fast \mm{f} changes in spacelike
directions. Hence, \mm{\nabla} is the spatial derivative, which proves
thus to be an absolute \p1{frame independent} operation, although absolute
space does not exist but spaces depend on observers. The explanation is
that spacelikeness is absolute and this is enough for spatial
derivative.

The derivative of a four-vector field \mmm{ \VVf: \VM \to \VVM } is
\mmm{\VVM \diad \VVM^*} valued at every point, the trace of which is the
four-divergence. In parallel, the spatial derivative \mmm{\VVq \diad
\nabla} of a three-vector field or spacelike vector field \mmm{ \VVq:
\VM \to \VVS } is \mm{ \VVS \diad \VVS^* } valued, the trace of which
is the three-divergence. On the other side, the spatial derivative of an
\mm{ \VM \to \VVS^* } three-covector field is \mm{ \VVS^* \diad \VVS^*
}, whose the antisymmetric part leads to the curl, \mm{\nabla \times\,,}
of the three-covector field.

An important special case is the derivative of a four-velocity field
\mmm{ \VVv: \VM \to V(1) \subset \VVM \Per \VVVT \,.} Since the
difference of any two elements of \mm{V(1)} is spacelike, differences of
four-velocities are always spacelike. Consequently, \mm{\VVv \diad \VVVD}
and \mm{\VVv \diad \nabla} are not only \mmm{\VVM \Per \VVVT \diad
\VVM^*} and \mm{\VVM \Per \VVVT \diad \VVS^*} valued, respectively, but
\mmm{\VVS \Per \VVVT \diad \VVM^*\,,} resp.\ \mm{\VVS \Per \VVVT \diad
\VVS^*} valued. Therefore, the three-divergence of a four-velocity
field is automatically meaningful, while the curl only after utilizing
the identification \mm{\VVS \equiv \VVS^* \diad \VVVL^{\VVVpow{2}}}
provided by \mm{\VVh \,.}

\end{document}